\documentclass[prd,superscriptaddress,nofootinbib,notitlepage]{revtex4-1}

\usepackage{epsfig,amsmath,amssymb,slashed}
\usepackage{graphicx}
\usepackage{color}
\usepackage{siunitx}
\usepackage{color}
\usepackage{bm}
\usepackage{subfig}
\usepackage[utf8]{inputenc}
\usepackage[english]{babel}
\usepackage[T1]{fontenc}
\usepackage{amsfonts}
\usepackage{amsthm}
\usepackage{hyperref}
\usepackage{wasysym}
\usepackage{xcolor}

\newcommand{\bra}[1]{\langle #1|}
\newcommand{\ket}[1]{|#1\rangle}

\newcommand{\media}[1]{\langle #1 \rangle}
\newcommand{\wad}[1]{\widehat a^\dagger(#1)}
\newcommand{\wa}[1]{\widehat a(#1)}
\newcommand{\wbd}{\widehat b^{\dagger}}
\newcommand{\wb}{\widehat b}
\newcommand{\wxi}{\widehat \xi}
\newcommand{\wxid}{\widehat \xi^{\dagger}}
\newcommand{\di}{{\rm d}}
\newcommand{\Tr}{{\rm Tr}}
\newcommand{\ii}{i}

\def\wT{{\widehat T}}
\def\wj{{\widehat j}}

\def\wPi{{\widehat{\Pi}}}

\def\wpsi{{\widehat{\psi}}}
\def\wrho{{\widehat{\rho}}}
\def\wrhol{{\widehat{\rho}_{\rm LE}}}

\def\wU{{\widehat{U}}}

\newcommand{\tr}{{\rm tr}}  
  
\newcommand{\e}{{\rm e}}

\newcommand{\p}{{\rm p}}

\newcommand{\be}{\begin{equation}}
\newcommand{\ee}{\end{equation}}                                                                               
\newcommand{\bea}{\begin{eqnarray}}
\newcommand{\eea}{\end{eqnarray}}

\newcommand{\wh}{\widehat}
\newcommand{\mT}{m_{\rm T}}

\newcommand{\pT}{{\bf p}_{\rm T}}

\begin{document}

\title{Relativistic quantum fluid with boost invariance} 

\author{D.\ Rindori}
\affiliation{Universit\`a di Firenze and INFN Sezione di Firenze\\ Via G.\ Sansone 1, Sesto Fiorentino, I-50019 Florence, Italy}
\author{L.\ Tinti}
\affiliation{Institut f\"ur Theoretische Physik, Johann Wolfgang Goethe-Universit\"at\\ Max-von-Laue-Str.\ 1, D-60438 Frankfurt am Main, Germany}
\affiliation{Instytut Fizyki,
Uniwersytet Jana Kochanowskiego w Kielcach\\
ul. Uniwersytecka \ 7,
PL 25-406 Kielce, Poland}
\author{F.\ Becattini}
\affiliation{Universit\`a di Firenze and INFN Sezione di Firenze\\ Via G.\ Sansone 1, Sesto Fiorentino, I-50019 Florence, Italy}
\author{D.H.\ Rischke}
\affiliation{Institut f\"ur Theoretische Physik, Johann Wolfgang Goethe-Universit\"at\\ Max-von-Laue-Str.\ 1, D-60438 Frankfurt am Main, Germany}
\affiliation{Helmholtz Research Academy Hesse for FAIR, Campus Riedberg, Max-von-Laue-Str.\ 12, D-60438 Frankfurt am Main, Germany}

\begin{abstract}
We study a relativistic fluid with longitudinal boost invariance in a quantum-statistical 
framework as an example of a solvable non-equilibrium problem. For the free quantum field, we calculate 
the exact form of the expectation values of the stress-energy tensor and the entropy current. For the
stress-energy tensor, we find that a finite value can be obtained only by subtracting the vacuum 
of the density operator at some fixed proper time $\tau_0$. As a consequence, the stress-energy 
tensor acquires non-trivial quantum corrections to the classical free-streaming form.  
\end{abstract}

\maketitle

\section{Introduction}

Spurred by a successful description of experimental data in high-energy nuclear collisions, relativistic hydrodynamics 
has recently made major progress, both regarding its theoretical foundations as well as its phenomenological applications. 
Lately, the quantum-statistical foundations of relativistic hydrodynamics have attracted a great deal of 
attention \cite{betaframe,hayata,kaminski,tinti,calabrese}, in particular to describe quantum phenomena in relativistic 
fluids such as chirality \cite{chirality} and polarization \cite{polarization}. In a quantum-statistical framework, hydrodynamic 
quantities, such as the stress-energy tensor and conserved currents, are the expectation values of the corresponding quantum operators with respect to a suitable statistical (or density) operator $\wrho$:
\be\label{meanset}
  T^{\mu\nu} = \tr (\wrho\,  \wT^{\mu\nu})_{\rm ren}\;,
\ee
where the subscript ``${\rm ren}$'' implies renormalization of the otherwise divergent expectation value.

In general, the form of the stress-energy tensor and the currents crucially depends on the density operator. Exact expressions are known only in a few 
cases, including the familiar global thermodynamic equilibrium and, as a recent development, global thermodynamic equilibrium 
with rotation and acceleration. However, no exact form is known in local thermodynamic equilibrium, 
which is defined by \cite{zubarev,weert,betaframe,hayata}:
\be\label{localdo}
 \wrho_{\rm LE} = \frac{1}{Z} \exp\left[ -\int_\Sigma \di \Sigma_\mu \left(\wT^{\mu\nu} \beta_\nu - 
 \zeta \wj^\mu \right) \right]\;,
\ee
where $\beta(x)$ is a four-temperature field [equal to the four-velocity $u(x)$ divided by the temperature $T(x)$], 
$\zeta(x)$ is a scalar field [equal to the ratio of the chemical potential $\mu(x)$ associated with the conserved 
current $\wj$ and the temperature]. The hypersurface $\Sigma$ is a three-dimensional space-like hypersurface, on 
which local equilibrium is defined. The calculation of expectation values of operators using Eq.\ \eqref{localdo} 
can be performed only in the hydrodynamic limit of slowly varying fields \cite{betaframe}. For the stress-energy 
tensor, the leading-order term coincides with the familiar perfect-fluid expression. Beyond this approximation, 
quantum corrections appear, which have been estimated by means of a perturbative expansion only in the global-equilibrium case \cite{becaquant}.

\begin{figure}
	\includegraphics[scale=0.85]{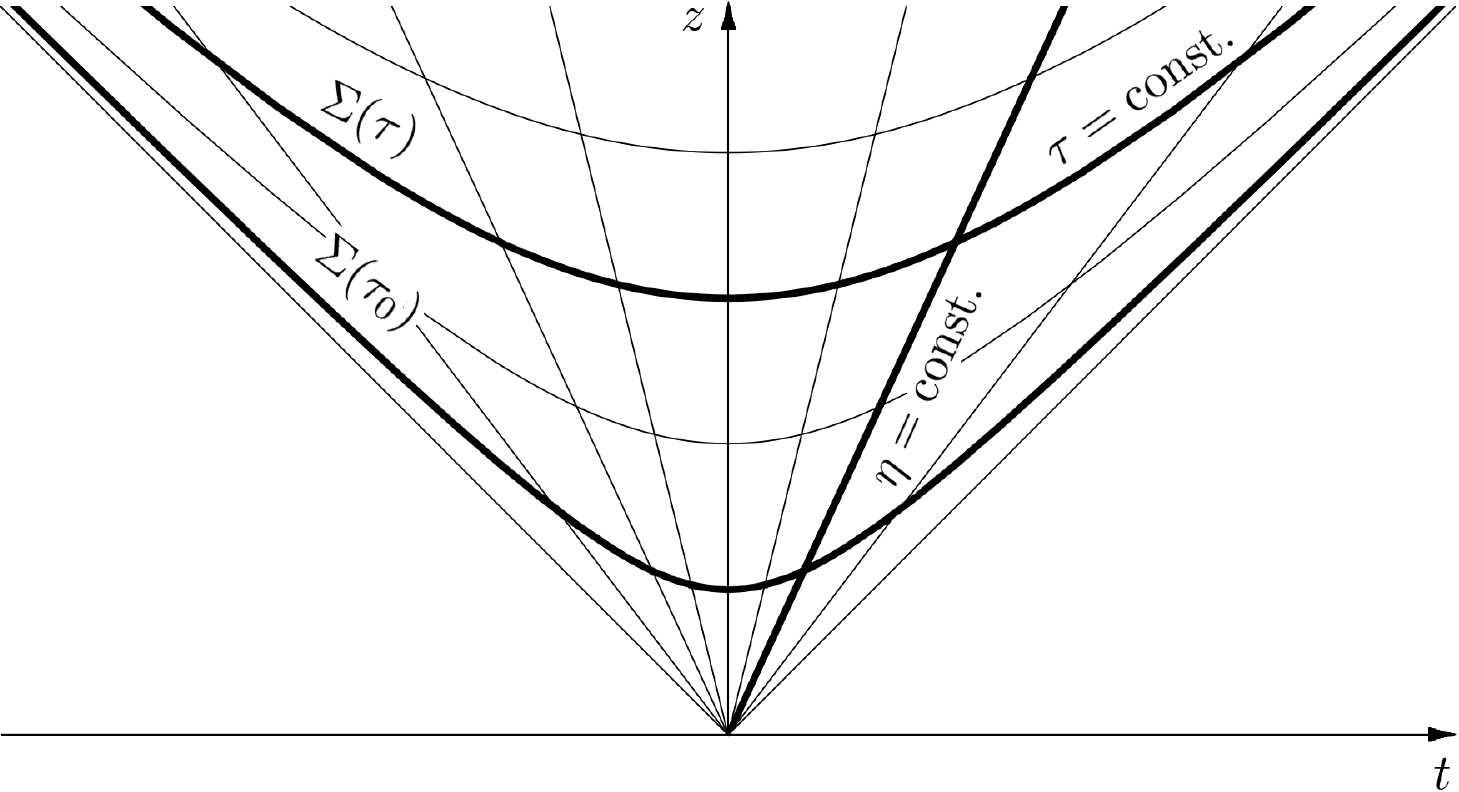}
	\caption{Two-dimensional section of the future light cone. Curves of constant Milne time 
	$\tau$ are hyperbolae, while curves of constant space-time rapidity $\eta$ are lines through the origin.
	The thicker hyperbolae are two-dimensional sections of the three-dimensional hypersurfaces $\Sigma(\tau)$ 
	and $\Sigma(\tau_0)$ at constant $\tau$ and constant $\tau_0$, respectively.}
	\label{figure}
\end{figure}

Recently, S.\ Akkelin \cite{Akkelin1,Akkelin2} has derived an exact solution of a particular non-equilibrium problem, a 
free neutral scalar field with the density operator:
\be\label{localdo_1}
 \wrho = \frac{1}{Z} \exp\left[ -\frac{1}{T(\tau)}\int_{\Sigma(\tau)} \di \Sigma_\mu \; \wT^{\mu\nu} u_\nu \right] \;, 
\ee
with $\Sigma(\tau)$ being a proper-time $\tau$ hyperbola in the future light-cone in two dimensions 
(see Fig.\ \ref{figure}) and $u(x)$ the four-velocity field coinciding with the unit vector perpendicular 
to $\Sigma$. The density operator \eqref{localdo_1} is invariant under longitudinal boosts, a symmetry 
which has been often used to study general features of relativistic hydrodynamics problems. Lately, 
longitudinal boost-invariant solutions have been studied in the context of spin-hydrodynamics \cite{florkowski} 
and magneto-hydrodynamics \cite{rischke1,rischke2,qunwang}.
This symmetry and the solution found in Ref.~\cite{Akkelin1} also offers a special opportunity to explore in 
detail some essential features of quantum relativistic hydrodynamics in a non-equilibrium situation and, in 
particular, to determine the pure quantum corrections to classical hydrodynamics and kinetic equations, including 
those to the stress-energy tensor and to the entropy current. In other words, this solution provides a benchmark 
test of a relativistic quantum fluid. 

In this work, we extend the results of Ref.~\cite{Akkelin1} and study the stress-energy 
tensor with longitudinal boost-invariant symmetry. We find that, even for the simplest case of a free 
scalar field, there are relevant quantum corrections related to its renormalization by subtraction 
of the vacuum expectation value. Indeed, while the traditional vacuum of the field, expanded in plane 
waves, the so-called Minkowski vacuum, fails to provide a finite energy density, the subtraction of the vacuum
expectation value with respect to the vacuum of the density operator does. Conversely, for the entropy current, 
no significant quantum correction is found.

This paper is organized as follows.
We will start in Sec.\ \ref{symmetry} with a review of the density-operator approach in relativistic 
quantum-statistical mechanics with special emphasis on symmetry considerations.
In Sec.\ \ref{quantum} we will specialize to the symmetry of concern for this work, that is boost invariance.
As underlying quantum field theory, in Sec.\ \ref{freefield} we will present the field theory of the free neutral 
scalar field in the future light cone, including a diagonalization of the density operator.
This will put us in the position to calculate the thermal expectation value of the stress-energy tensor in 
Sec.\ \ref{setensor} both in local thermodynamic equilibrium and out of equilibrium.
Finally, we will discuss the entropy current and entropy production in Sec.\ \ref{entropycurrent}, before concluding this paper in Sec.\ \ref{conclusions}.


In this work, we use natural units $\hbar=c=k_{\rm B}=1$. Operators in Hilbert space are denoted with a 
wide upper hat, e.g., $\widehat{O}$, while vectors of unit length have a regular hat, that is $\hat{n}^{\mu}$.
Repeated indices are assumed to be contracted. We adopt the ``mostly-minus'' convention, so the Minkowski metric 
is $g_{\mu \nu}={\rm diag}(1,-1,-1,-1)$. For the Levi-Civita symbol we use the convention $\epsilon^{0123}=1$.

\section{Local thermodynamic equilibrium, density operator, and symmetries}
\label{symmetry}

In quantum-statistical mechanics, the local-equilibrium density operator (LEDO) $\wrhol$, Eq.\  
\eqref{localdo}, is obtained by maximizing the entropy $S = -\Tr (\widehat{\rho}_{\rm LE} \log \widehat{\rho}_{\rm LE})$ 
under the constraints of fixed energy-momentum and, possibly, charge densities on a given three-dimensional 
space-like hypersurface $\Sigma$. The hypersurface can be either specified {\it a priori} or can be found 
in a self-consistent procedure by using the thermodynamic fields themselves \cite{betaframe}.

The energy-momentum densities on a hypersurface $\Sigma$ are obtained by contracting the stress-energy tensor
with its normal unit vector $n$, so that the constraints read:
\begin{equation}\label{constr}
    n_\mu \Tr(\wrhol \,\wT^{\mu \nu})_{\rm ren}= n_{\mu} T^{\mu \nu},\qquad
\end{equation}
and likewise for the conserved currents. The densities on the right-hand side of Eq.\ \eqref{constr} are meant to be the 
actual ones, no matter how they are known or defined, and they are supposedly finite. It is crucial to specify 
that the expectation values on the left-hand side must be suitably renormalized because, in general, the expectation 
value of the operator $\wT^{\mu\nu}$ with a density operator such as in Eq.\  \eqref{localdo} is divergent. For 
instance, in free field theory, the renormalization procedure is most readily established by subtracting 
the vacuum expectation value, that is:
\begin{equation}\label{vacsubtr}
        \Tr(\wrho \,\wT^{\mu \nu})_{\rm ren}
        =\Tr(\wrho \,\wT^{\mu \nu})-\langle 0|\wT^{\mu \nu}|0\rangle\;, 
\end{equation}
which is tantamount to normal-ordering of the creation and annihilation operators because the 
currents are quadratic in the fields. We will delve into the question of vacuum subtraction in Sec.~\ref{vacuum}.

With the constraints \eqref{constr}, the function to be maximized with respect to $\widehat{\rho}_{\rm LE}$ 
is:
\begin{equation}
    -\Tr (\wrhol \log \wrhol) +\int_{\Sigma(\tau)} \di\Sigma_\mu \;
    \left[ \left( \tr(\wrhol \wT^{\mu \nu})_{\rm ren} -T^{\mu \nu}\right) \beta_{\nu}-
    \zeta\left( \tr(\wrhol \wj^{\mu})_{\rm ren} -j^{\mu} \right)\right]\;,
\end{equation}
where the thermodynamic fields $\beta$ and $\zeta$ are Lagrange multipliers introduced to 
enforce the constraints \eqref{constr}. The solution is Eq.\ \eqref{localdo} and
it should be pointed out that it can be kept in that simple form without subtraction of the vacuum
expectation value because the latter is not an operator and would appear in the partition
function $Z$ as well (in order to make $\Tr \wrhol = 1$), hence cancelling out in the ratio. 
With the energy-momentum densities given by the right-hand side 
of Eq.\ \eqref{constr}, the thermodynamic fields $\beta$ and $\zeta$ are determined by 
solving them with $\wrhol$ given by Eq.\ \eqref{localdo}; there are five equations with five unknowns ($\beta^\mu$
and $\zeta$), which in general can be solved.

Unless $\beta$ is a Killing field and $\zeta$ constant, which characterizes a state of global thermodynamic
equilibrium, the operator \eqref{localdo} is not independent of the hypersurface, hence it cannot 
be the actual density operator in the Heisenberg representation. In fact, the true density operator
is, for a system which supposedly achieves local thermodynamic equilibrium at some time $\tau_0$, 
the so-called non-equilibrium density operator (NEDO), which is just Eq.\ \eqref{localdo} at time $\tau_0$:
\be\label{nedo}
 \wrho = \frac{1}{Z} \exp\left[ -\int_{\Sigma(\tau_0)} \di \Sigma_\mu \left( \wT^{\mu\nu} \beta_\nu - 
 \zeta \wj^\mu \right) \right]\;.
\ee
This can be recast by using Gauss' theorem as \cite{becazuba}
\be\label{nedo2}
 \wrho =  \dfrac{1}{Z} 
 \exp\left[ - \int_{\Sigma(\tau_0)} \!\!\!\!\!\! \di \Sigma_\mu \; \left( \wT^{\mu\nu} 
  \beta_\nu - \wj^\mu \zeta \right) \right] =
 \dfrac{1}{Z} 
 \exp\left[ - \int_{\Sigma(\tau)} \!\!\!\!\!\! \di \Sigma_\mu \; \left( \wT^{\mu\nu} 
  \beta_\nu - \wj^\mu \zeta \right) + \int_\Omega \di \Omega \; \left( \wT^{\mu\nu} 
  \nabla_\mu \beta_\nu - \wj^\mu \nabla_\mu \zeta \right) \right] \; .
\ee
In the exponent on the right-hand side, the first term is just the operator of local equilibrium
at time $\tau$, while the second term contains dissipative corrections \cite{becazuba}.

Suppose now that the actual density operator, the NEDO, has some symmetry, meaning that it 
commutes with some unitary representation $\wU(g)$ in Hilbert space of a group or a subgroup 
G of transformations, to be specific of the proper orthochronous Poincar\'e group IO(1,3)$^\uparrow_+$.
We have:
\begin{align*}
 \wU(g) \wrho \, \wU(g)^{-1} &= \frac{1}{Z} \exp\left[ -\int_{\Sigma(\tau_0)} \di \Sigma_\mu(x) \left( 
  \wU(g) \wT^{\mu\nu}(x) \wU(g)^{-1} \beta_\nu(x) -\zeta(x) \wU(g) \wj^\mu(x) \wU(g)^{-1} \right) \right]
 \\
 &= \frac{1}{Z} \exp\left[ -\int_{\Sigma(\tau_0)} \di \Sigma_\mu(x) \left( 
  D(g^{-1})^\mu_\rho D(g^{-1})^\nu_\sigma \wT^{\rho\sigma}(g(x)) \beta_\nu(x) -
   \zeta(x) D(g^{-1})^\mu_\rho \wj^\rho(g(x)) \right) \right]\;.
\end{align*}
Let us now set $y=g(x)$ and we obtain, remembering $\di \Sigma_\mu (x) = D(g)^\nu_\mu \di \Sigma_\nu(y)$,
$$
 \wU(g) \wrho \, \wU(g)^{-1} = \frac{1}{Z} \exp\left[ -\int_{g(\Sigma(\tau_0))} \di \Sigma_\rho(y) \left( 
  \wT^{\rho\sigma}(y) D(g^{-1})^\nu_\sigma \beta_\nu(g^{-1}(y)) - \zeta(g^{-1}(y)) \wj^\rho(y) \right)
   \right].
$$
Thus, if the hypersurface is invariant under the transformation $g$ and if:
\be\label{therminv}
   D(g^{-1})^\nu_\sigma \beta_\nu(g^{-1}(y)) = \beta_\sigma(y)\;,
   \qquad \zeta(g^{-1}(y)) = \zeta(y)\;,
\ee
then the operator $\wrho$ is invariant under the transformation $\wU(g) \wrho \wU(g)^{-1}$. 
Equations \eqref{therminv} specify the symmetry conditions on the transformations of the 
thermodynamic fields $\beta$ and $\zeta$. 
An invariance of $\wrho$ has straightforward consequences for the expectation values of operators.
For instance, for the stress-energy tensor:
\be\label{tensor}
  T^{\mu\nu}(x) = \Tr [\wrho \wT^{\mu\nu}(x)] = \Tr \left[\wrho \, \wU(g)^{-1} \wT^{\mu\nu}(x) \wU(g)\right]
  = D(g)^\mu_\rho D(g)^\nu_\sigma\, \Tr \left[\wrho\, \wT^{\mu\nu}(g^{-1}(x))\right] =
   D(g)^\mu_\rho D(g)^\nu_\sigma\, T^{\mu\nu} (g^{-1}(x))\; .
\ee
If we consider a one-parameter subgroup of transformations $g_\phi$ [e.g., a rotation, $g_\phi=\exp (-\ii \phi {\sf J})$, around 
some axis], Eqs.\ \eqref{therminv} and \eqref{tensor}
have the consequence that the Lie derivative along the vector field $X(x) = \di g_\phi(x) / \di \phi$ 
of the field under consideration vanishes, that is:
\be\label{lie}
  {\cal L}_X (\beta)^\mu = 0 \;,\qquad \qquad {\cal L}_X (T)^{\mu\nu} = 0 \; .
\ee

An important question concerns the persistence of the symmetry of the local thermodynamic
equilibrium operator, that is whether the implication:
$$
  \wrho = \wU(g) \wrho \, \wU(g)^{-1}  \implies \wrhol(\tau) = \wU(g) \wrhol(\tau) \, \wU(g)^{-1} 
$$
is true for any $\tau$. Indeed, it can be shown that if the subgroup G transform $\Sigma(\tau_0)$ 
into  itself and if the fields $\beta$ and $\zeta$ are also symmetric under G, namely they 
fulfill Eqs.\ \eqref{therminv} or \eqref{lie}, this is the case. Indeed, by definition, $\wrhol(\tau)$ is the 
solution of maximizing a function which is invariant under any unitary transformation, the entropy, 
with the constraint \eqref{constr}. If a particular $\wrhol$ fulfills Eq.\ \eqref{constr}, so 
will $\wU(g) \wrhol(\tau) \wU(g)^{-1}$ as it can be readily checked. Therefore, either 
$\wU(g) \wrhol(\tau) \wU(g)^{-1}$ is a different solution of the constrained maximization problem, 
or it coincides with $\wrhol(\tau)$. In both cases, it is possible to generate one symmetric solution 
under the subgroup G by using a particular solution $\wrhol^{(0)}$ and summing over all $g$'s: 
$$
  \wrhol(\tau) = \frac{1}{M(G)} \sum_{g \in G} \wU(g) \wrhol^{(0)} \wU(g)^{-1}\; .
$$
It is then obvious that the sufficient condition for $\wrhol(\tau)$, given by Eq.\ \eqref{localdo}, 
to be symmetric under G is that the fields $\beta$ and $\zeta$ fulfill Eqs.\ \eqref{therminv}
at time $\tau$. This is a crucial point for the purpose of this work.

\section{Relativistic quantum fluid with longitudinal boost invariance}
\label{quantum}

Suppose that the density operator is given by Eq.\ \eqref{nedo} with $\Sigma(\tau_0)$
being the hyperboloid $\tau = \sqrt{t^2 - z^2} = \tau_0$ in Minkowski space-time and with 
\be\label{beta}
  \beta^\mu = \frac{1}{T(\tau_0)} \frac{1}{\tau_0} (t,0,0,z) = \frac{1}{T(\tau_0)} u^\mu\;,
\ee
where $T(\tau_0)$ and $\zeta(\tau_0)$ are constant 
on the hypersurface. This vector field is time-like on the hypersurface $\Sigma(\tau_0)$, 
hence thermodynamically meaningful.

The field $\beta$ in Eq.~\eqref{beta} and the field $\zeta$ fulfill Eq.\ \eqref{therminv}
for any longitudinal boost with hyperbolic angle $\xi$ along the $z$ axis, ${\sf L}_z(\xi)$,
and manifestly for translations and rotations in the $xy$ plane. Besides, the hypersurface
$\Sigma(\tau_0)$ is invariant under the same transformations. Therefore, the density operator 
has the symmetry group ${\rm IO}(2) \otimes {\rm SO}(1,1)$, that is the Euclidean group in
the transverse plane times Lorentz transformations in the longitudinal direction. Furthermore, the density operator is also 
invariant under a space-reflection transformation turning $x,y,z$ into $-x,-y,-z$. 

This symmetry group dictates the possible forms of vector and tensor fields, which are most
easily found by using Milne coordinates, $(\tau,x,y,\eta)$, instead of the usual Cartesian ones,
$(t,x,y,z)$:
\begin{align*}
  &    t=\tau \cosh \eta\;,\qquad z=\tau \sinh \eta\;, \\
  &    \tau = \sqrt{t^2-z^2}\;, \qquad \eta = \frac{1}{2} \log \left( \frac{t+z}{t-z}\right)\;,
\end{align*}
whence it turns out that the coordinate basis vectors are:
\begin{align*}
  \frac{\partial}{\partial \tau} = \frac{1}{\tau}(t,0,0,z)= (\cosh \eta,0,0,\sinh \eta) = u\;,
  \qquad &\frac{\partial}{\partial \eta} = \tau (\sinh \eta,0,0,\cosh \eta) = (z,0,0,t)
  \equiv \tau \hat\eta\;, \\
  \frac{\partial}{\partial x} = \hat i\;, \qquad & \frac{\partial}{\partial y} = \hat j\;,
\end{align*}
and the metric tensor:
$$
    {\rm d}s^2={\rm d}t^2-{\rm d}x^2-{\rm d}y^2-{\rm d}z^2=
    {\rm d}\tau^2-{\rm d}x^2-{\rm d}y^2-\tau^2\,{\rm d}\eta^2\;.
$$
The vector fields $X(x)$ associated with the symmetry group along which the Lie derivatives 
vanish can be readily found:
\begin{align}\label{liefields}
 \frac{\di {\sf T_1}(a)x}{\di a} = \hat i\;, \qquad & \frac{\di {\sf T_2}(a)y}{\di a} = \hat j\;, \\
 \nonumber
 \frac{\di {\sf R}(\phi)x}{\di \phi} = (0,-y,x,0) \equiv r \hat \varphi\;, \qquad & 
 \frac{\di {\sf L}_3(\xi)x}{\di \xi} = (z,0,0,t) = \tau \hat\eta\;, 
\end{align}
where ${\sf T}_i$ are translations in the coordinate directions of the $xy$ plane, $r=\sqrt{x^2+y^2}$,
${\sf R}(\varphi)$ is a rotation with angle $\varphi$ in the same plane, and ${\sf L}_3(\xi)$ is a 
longitudinal boost with hyperbolic angle $\xi$. Note that three vector fields are just the Milne-coordinate basis vectors, which, by construction, have vanishing Lie derivatives among each other,
that is vanishing Lie commutators. 

As has
been mentioned, the condition of vanishing Lie derivatives along the vector fields \eqref{liefields} puts strong limitations on the form of the fields in general. For instance, a vector field 
$V(x)$ can be decomposed onto the coordinate basis vectors:
$$
  V(x) = A(\tau) u + B(\tau) \hat i + C (\tau) \hat j + D(\tau) \tau \hat \eta\;,
$$
where the coefficients depend on the variable $\tau$ only as a consequence of ${\cal L}_X(V)=0$, 
where $X$ is either $\hat i$, or $\hat j$, or $\tau \hat \eta$. Also, by implementing ${\cal L}_{\hat \varphi}(V) = 0$
one obtains that both $B$ and $C$ are in fact zero. Furthermore, by reflection invariance, the component 
proportional to $\hat\eta$ must be vanishing because a reflection turns $\eta$ into $-\eta$ and the 
vector field has just one component:
\be\label{vecfield}
   V(x) = A(\tau) u.
\ee
Similarly, the form a symmetric tensor field like the stress-energy tensor $T^{\mu\nu}$ can 
be obtained by iterated projections onto vectors and orthogonal components. The result is:
\be\label{set}
    T^{\mu \nu} = {\cal E}(\tau)u^{\mu}u^{\nu}+ {\cal P}_{\rm T}(\tau) \left(
     \hat i^{\mu}\hat i^{\nu}+\hat j^{\mu} \hat j^{\nu} \right) + {\cal P}_{\rm L}(\tau)\hat \eta^{\mu} 
      \hat \eta^{\nu}.
\ee
The form \eqref{set} is different from the usual perfect fluid form, for which ${\cal P}_{\rm T}={\cal P}_{\rm L}$.
The difference between transverse and longitudinal pressure is owing to the lack of full rotational symmetry 
and it is to be expected, on general grounds, that this difference is a quantum effect, as already observed 
for global equilibrium \cite{becaquant}.

In order to determine the three scalar functions in Eq.\ \eqref{set}, we have to calculate the expectation 
values of operators with the density operator \eqref{nedo}. The unit four-vector orthogonal to the hyperboloid 
with fixed $\tau$ is $u$ itself, so the operator \eqref{nedo} becomes:
\begin{equation}\label{nedo3}
    \wrho = \frac{1}{Z}\exp \left[-\frac{\wPi(\tau_0)}{T(\tau_0)} \right]\;,
\end{equation}
with:
$$
  \wPi(\tau_0) = \int_{\Sigma(\tau_0)} \di \Sigma \; u_\mu u_\nu \wT^{\mu\nu} = 
  \tau_0 \int \di x \, \di y \, \di \eta \; u_\mu u_\nu \wT^{\mu\nu}\;,
$$   
where we have used the measure in Milne coordinates. It should be stressed that the operator
$\wPi(\tau_0)$ is not conserved because the divergence of the integrand is not zero:
$$
  \partial_\mu \left( u_\nu \wT^{\mu\nu} \right) = \wT^{\mu\nu} \partial_\mu u_\nu \ne 0\;,
$$
so it depends on $\tau_0$. We can also write down a general form of the local equilibrium operator 
$\wrhol(\tau)$ at any Milne time $\tau$ by taking the hyperboloid 
$\tau=const.$ as local-equilibrium hypersurface, which is invariant under the same transformations as $\Sigma(\tau_0)$, according to 
the discussion in Sec.~\ref{symmetry}. Since the field $\beta(\tau)$ must fulfill Eqs.\ \eqref{therminv} 
and $\eqref{lie}$, it can only be of the form \eqref{vecfield}:
$$
  \beta = \frac{1}{T(\tau)} u = \frac{1}{T(\tau)} (\cosh \eta, 0, 0, \sinh \eta)\;,
$$
thus the constraint \eqref{constr} becomes, by using Eq.\ \eqref{set}:
$$
   n_\mu \Tr(\wrhol \,\wT^{\mu \nu})_{\rm ren} \equiv n_\mu T^{\mu\nu}_{\rm LE} = 
   u_\mu T^{\mu\nu}_{\rm LE} = {\cal E}(\tau)_{\rm LE} u^\nu = n_\mu T^{\mu\nu} = 
   {\cal E}(\tau) u^\nu.
$$
This vector equation comes down to one scalar equation ${\cal E}(\tau)_{\rm LE} = {\cal E}(\tau)$ with $T(\tau)$ 
as unknown to be determined once the actual ${\cal E}(\tau)$ is determined by using the actual 
density operator \eqref{nedo}. The local thermodynamic equilibrium operator will be of the
same form as Eq.\ \eqref{nedo3}, that is:
\begin{equation}\label{ledo2}
    \wrhol(\tau) = \frac{1}{Z} \exp \left[ -\frac{\wPi(\tau)}{T(\tau)} \right]\;,
\end{equation}
with $\wPi(\tau) \ne \wPi(\tau_0)$.

\subsection{Vacuum effects}\label{vacuum}

A very interesting feature of a relativistic quantum fluid with the four-temperature field \eqref{beta} is 
that the spectrum of $\wPi(\tau_0)$, and particularly the lowest-lying eigenvector, the $\wPi$ vacuum, may 
depend on $\tau$, as it is clear from Ref.~\cite{Akkelin1}. This $\tau$-dependent vacuum $\ket{0_\tau}$ 
is in general also different from the vacuum of a quantum field theory -- even for free fields -- in flat space-time 
obtained by quantizing in Cartesian coordinates, the so-called Minkowski vacuum $\ket{0_M}$. This is clearly 
at variance with familiar equilibrium quantum thermodynamics, where the Hamiltonian operator achieves 
its minimal eigenvalue in the Minkowski vacuum. The distinction between vacua is very important as to the 
renormalization of several quantities, including, e.g., the stress-energy tensor. In a free field theory, the 
renormalization of the expectation value of an operator $\widehat O$ involves the subtraction of its vacuum expectation 
value. If more vacua are present, there is an ambiguity as we could define, as usual:
\be\label{minksub}
  \media{\widehat O}_{\rm ren} \equiv \Tr (\wrho \widehat O) - \bra{0_M} \widehat O \ket{0_M}\;,
\ee
[see Eq.~\eqref{vacsubtr}] or, in our case:
\be\label{tausub}
  \media{\widehat O}_{\rm ren} \equiv \Tr( \wrho \, \widehat O) - \bra{0_\tau} \widehat O \ket{0_\tau}\;.
\ee
Note that the $\wPi$ vacuum can be subtracted by taking the limit $T(\tau) \to 0$ of the unrenormalized expression 
since:
$$
 \lim_{T(\tau) \to 0} \wrhol(\tau) = \lim_{T(\tau) \to 0} \frac{1}{Z} \exp[-\wPi(\tau)/T(\tau)]
  = \ket{0_\tau} \bra{0_\tau} \equiv {\sf P}_{0_\tau}\;.
$$
For this reason, in general the vacuum $\ket{0_\tau}$ will have the same symmetries as the original 
density operator, but it will be less symmetric than the supposedly Poincar\'e-invariant Minkowski 
vacuum $\ket{0_M}$ \footnote{This does not mean that the vacuum $\ket{0_\tau}$ is degenerate, but that
Poincar\'e transformations will give rise to non-vanishing components of excited states.}. 

It should be pointed out that the vacuum $\ket{0_\tau}$ is $\tau$-dependent, hence a subtraction like in Eq.~\eqref{tausub} implies that the expectation value can get an undesired time dependence. For instance, if we define 
the renormalized stress-energy tensor as:
$$
  T^{\mu\nu} \equiv \Tr(\wrho \, \wT^{\mu\nu}) - \bra{0_\tau} \wT^{\mu\nu} \ket{0_\tau}
  = \Tr [ (\wrho - {\sf P}_{0_\tau}) \wT^{\mu\nu} ]\;,
$$
then:
$$
 \partial_\mu T^{\mu\nu} = \partial_\mu \Tr [ (\wrho - {\sf P}_{0_\tau}) \wT^{\mu\nu} ]
 = \Tr [ (\wrho - {\sf P}_{0_\tau}) \partial_\mu \wT^{\mu\nu} ] + 
 \Tr [- (\partial_\mu {\sf P}_{0_\tau}) \wT^{\mu\nu}] =  - \Tr \left[ 
 u_\mu \frac{\partial {\sf P}_{0_\tau}}{\partial \tau}  \wT^{\mu\nu} \right] \ne 0\;,
$$
where we used $\partial_\mu \wT^{\mu\nu} = 0$ and the time independence of the density operator.
Therefore, the expectation value $T^{\mu \nu}$ would no longer fulfill a conservation equation even though the operator 
$\wT$ does. 

Therefore, in order to have a properly finite, conserved stress-energy tensor for a  relativistic quantum
fluid, the vacuum must be necessarily fixed, just like the density operator. Of course the Minkowski
vacuum $\ket{0_M}$ meets this requirement and is seemingly the most obvious choice. However, we will
see in Sec.~\ref{setensor} that the subtraction of the vacuum expectation value of $\wT^{\mu\nu}$ of a 
free field with respect to $\ket{0_M}$
does not give rise to a finite value, for the particular symmetry we are dealing with, and 
an alternative definition is needed.

\section{Free scalar field in Milne coordinates} 
\label{freefield}

As has been mentioned in the Introduction, a closed analytic form of the stress-energy tensor 
with the four-temperature field \eqref{beta} exists for the case of free fields, providing
the opportunity to determine exact quantum corrections to the classical expressions in the 
non-equilibrium case. The system which is described by the operator \eqref{nedo} and a 
free scalar field is that of a fluid where interactions effectively cease at the hypersurface 
$\Sigma(\tau_0)$ with temperature $T(\tau_0)$ and a four-velocity $u= T \beta$, with particles 
freely streaming thereafter. We thus expect to recover, in the classical limit, the classical 
kinetic-theory solutions of the free-streaming Boltzmann equation starting from the local thermodynamic 
equilibrium expressions with proper temperature $T(\tau_0)$ and flow velocity $u(\tau_0)$.

The calculation of the stress-energy tensor for the massive free scalar field $\wpsi(x)$ requires the 
solution of the Klein-Gordon equation in Milne coordinates:
$$
    \left[ \frac{1}{\tau} \partial_\tau \left(\tau \partial_{\tau} \right) -\partial_x^2-\partial_y^2-
    \frac{1}{\tau^2}\partial_{\eta}^2+m^2 \right] \wpsi(\tau,{\bf x}_{\rm T},\eta)=0\;.    
$$
This is a well-known problem in the literature \cite{Padmanabhan,Arcuri}, which has even raised 
some discussion. It has been convincingly demonstrated \cite{Arcuri} that, within the future light cone, 
there is a complete set of solutions of the Klein-Gordon equation in Milne coordinates, which allow an expansion in terms of the 
familiar plane waves and which do not mix positive and negative frequencies. These mode functions can be 
obtained starting from the usual expansion of the scalar field \cite{Akkelin1} in plane waves. We
will recapitulate the salient points of the derivation presented in Ref.\ \cite{Akkelin1}. The obtained
full expansion of the field in Milne coordinates reads:
\begin{equation}\label{field}
    \wpsi(\tau,{\bf x}_{\rm T},\eta)=\int\frac{{\rm d}^2{\rm p_T}\,{\rm d}\mu}
    {4\pi \sqrt{2}}\left[h({\sf p},\tau){\rm e}^{i(\pT\cdot {\bf x}_{\rm T}+\mu\eta)}
    \wb_{\sf p}+h^*({\sf p},\tau){\rm e}^{-i(\pT\cdot {\bf x}_{\rm T}+\mu\eta)}
    \wbd_{\sf p}\right]\;,
\end{equation}
where ${\sf p}=(\pT,\mu)$ to distinguish it from the Cartesian vector ${\bf p}=(\pT,p_z)$. Here, $\wbd_{\sf p}$ 
and $\wb_{\sf p}$ are creation and annihilation operators satisfying the usual algebra:
\begin{equation}\label{commrel}
    [\wb_{\sf p},\wbd_{{\sf p}'}]=\delta^2(\pT-\pT')\delta(\mu-\mu')\;,\qquad
    [\wb_{\sf p},\wb_{{\sf p}'}]=0=[\wbd_{\sf p},\wbd_{{\sf p}'}]\;.
\end{equation}
The relation between the operators $\wbd_{\sf p}$ and the familiar $\widehat{a}^\dagger(p)$ 
of the plane-wave expansion reads:
\be\label{atob}
  \widehat{a}^\dagger(p) = \frac{1}{\sqrt{2\pi \mT \cosh y}} 
  \int_{-\infty}^{+\infty}\di \mu \; \e^{-\ii \mu y} \, \wbd_{{\sf p}} \;,
\ee
where $y$ is the particle rapidity
in longitudinal direction, which can be easily inverted to obtain
$p_z$. Since there is no mixing between creation 
and annihilation operators, the vacuum of the $b_{{\sf p}}$ operators is the same Minkowski vacuum $\ket{0_M}$
as for the operators $a(p)$, which is a consequence of the fact that the functions $h({\sf p},\tau)$ can be
expressed as a linear combination of plane waves with just {\em positive} frequency \cite{winitzki}. 
In Eq.\  \eqref{field} $\mu$ is the eigenvalue of the boost operator $\widehat K_z$, so that:
$$
 \wh{U}({\sf L}_3(\xi)) \wbd_{{\sf p}} \wh{U}({\sf L}_3(\xi))^{-1} = 
 \e^{-\ii \xi \widehat K_z} \wbd_{{\sf p}} \e^{\ii \xi \widehat K_z} = 
 \e^{-\ii \xi \mu} \wbd_{{\sf p}}\;,
$$
i.e., $\wbd_{\sf p}$ creates a state with eigenvalue $\mu$. The $\tau$-dependent functions 
in Eq.\ \eqref{field} are
\begin{equation}\label{hankel}
    h({\sf p},\tau)=-\ii {\rm e}^{\frac{\pi}{2} \mu}{\rm H}^{(2)}_{\ii \mu}(m_{\rm T}\tau)\;,\qquad
    h^*({\sf p},\tau)=\ii {\rm e}^{-\frac{\pi}{2} \mu}{\rm H}^{(1)}_{\ii \mu}(m_{\rm T}\tau)\;,
\end{equation}
where the Hankel functions are \cite{Gradshteyn}:
\begin{equation}\label{intrepr}
    \begin{split}
 {\rm H}^{(2)}_{\ii \mu}(m_{\rm T}\tau)=&-\frac{1}{\ii \pi}{\rm e}^{-\frac{\pi}{2}\mu}
 \int_{-\infty}^{+\infty}{\rm d}\theta \,{\rm e}^{-\ii m_{\rm T}\tau \cosh \theta +\ii \mu \theta}\;,\\
 {\rm H}^{(1)}_{\ii \mu}(m_{\rm T}\tau)=&\frac{1}{\ii \pi}{\rm e}^{\frac{\pi}{2} \mu} 
 \int_{-\infty}^{+\infty}{\rm d}\theta \,{\rm e}^{\ii m_{\rm T}\tau \cosh \theta- \ii \mu \theta}\;,
    \end{split}
\end{equation}
with $m_{\rm T}= \sqrt{\p_{\rm T}^2+m^2}$ being the transverse mass. The integration variable $\theta$
in Eq.\  \eqref{intrepr} is related to the Milne coordinates and rapidity by \cite{Akkelin1}:
\be\label{etashift}
   \theta = y - \eta\;.
\ee
The functions \eqref{intrepr} solve the differential equations:
$$
 \left[ \frac{1}{\tau} \partial_\tau \left(\tau \partial_{\tau} \right) + m_T^2 + \frac{\mu^2}{\tau^2} 
  \right] h({\sf p},\tau) = 0\;,
$$
which are indeed Bessel's differential equations. It is also useful to define:
\be\label{pitau}
  \omega^2 = m_{\rm T}^2+\frac{\mu^2}{\tau^2}\;.
\ee

Let us now work out the density operator, particularly the operator $\wPi(\tau)$ in 
Eq.\ \eqref{ledo2}. In a non-equilibrium situation it is known that the density operator depends 
on the particular stress-energy tensor operator which is employed, however for the free scalar field
we will be using the canonical tensor:
$$
    \wT_C^{\mu \nu}=\frac{1}{2}\left(\partial^{\mu} \,\wpsi\partial^{\nu}\wpsi + 
    \partial^{\nu}\wpsi \, \partial^{\mu}\wpsi\right)-g^{\mu \nu}\widehat{\cal L}\; ,\qquad
    \widehat{\cal L}=\frac{1}{2}\left(g^{\mu \nu}\partial_{\mu} \, \wpsi\partial_{\nu}\wpsi-m^2\wpsi^2\right)\;,
$$
where $\widehat{\cal L}$ is the Lagrangian density. Hence:
\be\label{setmilne}
    \wT_C^{\mu \nu}u_{\mu}u_{\nu}=
    \frac{1}{2}\left[\left(\partial_{\tau}\wpsi\right)^2+\left(\partial_x\wpsi\right)^2+
    \left(\partial_y\wpsi\right)^2+\frac{1}{\tau^2}\left(\partial_{\eta}\wpsi\right)^2+m^2\wpsi^2\right]\;.
\ee
By using the above equation along with Eq.\ \eqref{field} and taking advantage of the invariance by 
reflection ${\sf p} \to -{\sf p}$ of the functions $h({\sf p},\tau)$, one can obtain the following 
expression for $\wPi(\tau)$:
\begin{equation}\label{Pitau}
        \wPi(\tau) = \tau \int \di x \, \di y \, \di \eta \; \widehat{T}^{\mu \nu}u_{\mu}u_{\nu} = 
        \int{\rm d}^2{\rm p_T}\,{\rm d}\mu\,\frac{\omega}{2}\left[
    K\left(\wb_{\sf p} \wbd_{\sf p} + \wbd_{\sf p} \wb_{\sf p} \right)+
    \Lambda \wb_{\sf p}\wb_{-{\sf p}}+\Lambda^*\wbd_{\sf p}\wbd_{-{\sf p}}
    \right]\;,
\end{equation}
where the positive real function $K({\sf p},\tau)$ and the complex function $\Lambda({\sf p},\tau)$ 
are defined as:
\begin{equation}\label{kappa}
    K({\sf p},\tau)=\frac{\pi \tau}{4\omega}\left(\left|\partial_{\tau}h({\sf p},\tau)\right|^2+
    \omega^2 |h({\sf p},\tau)|^2\right)\;,
\end{equation}
\begin{equation}\label{lambda}
    \Lambda({\sf p},\tau)=\frac{\pi \tau}{4\omega}\left\{\left[\partial_{\tau}h({\sf p},\tau)\right]^2+
    \omega^2 h^2({\sf p},\tau)\right\}\;.
\end{equation}
Note that, with $\omega$ and $h$ being invariant under a reflection ${\sf p} \to -{\sf p}$, so 
are $K$ and $\Lambda$, and:
\begin{equation}\label{relkl}
    K^2({\sf p},\tau)-|\Lambda({\sf p},\tau)|^2=1\;,
\end{equation}
as $K^2-|\Lambda|^2$ is proportional to the Wronskian of the Hankel functions
$$
    K^2({\sf p},\tau)-|\Lambda({\sf p},\tau)|^2=-\left(\frac{\pi m_{\rm T}\tau}{4}\right)^2\left(W[{\rm H}^{(2)}_{i\mu}(m_{\rm T}\tau),{\rm H}^{(1)}_{i\mu}(m_{\rm T}\tau)]\right)^2\;,
$$
which is known to be a very simple function \cite{Gradshteyn}:
\be\label{wronsk}
    W[{\rm H}^{(2)}_{i\nu}(x),{\rm H}^{(1)}_{i\nu}(x)]={\rm H}^{(2)'}_{i\nu}(x){\rm H}^{(1)}_{i\nu}(x)-{\rm H}^{(1)'}_{i\nu}(x){\rm H}^{(2)}_{i\nu}(x)=\frac{4i}{\pi x}\;.
\ee
The above relation is not accidental but it is related to the invariance of the Klein-Gordon scalar 
product of the mode functions \cite{winitzki}. Equation \eqref{relkl} allows to write:
\begin{equation}\label{reltheta}
    K({\sf p},\tau) = \cosh 2 \Theta ({\sf p},\tau)\;, \qquad \qquad \Lambda({\sf p},\tau) = \sinh 2 \Theta ({\sf p},\tau)
    \exp[ \ii \chi ({\sf p},\tau)]\;,
\end{equation}
which is very important to highlight the vacuum effects, as it will become clear later. 

Due to the terms proportional to $\Lambda$ and $\Lambda^*$, $\wPi(\tau)$ in Eq.\ \eqref{Pitau} is not 
diagonal in the creation and annihilation operators. If it were, we could easily calculate the 
expectation values of products of creation and annihilation operators, hence of operators quadratic 
in the field, using standard methods. We thus look for a suitable Bogolyubov transformation that 
diagonalizes $\wPi(\tau)$,
\begin{equation}\label{bogo1}
    \begin{split}
        &\wxid_{\sf p}(\tau)=A({\sf p}, \tau)\wbd_{\sf p}-B({\sf p},\tau)\wb_{-{\sf p}}\;,\\
        &\wxi_{\sf p}(\tau)=A^*({\sf p},\tau)\wb_{\sf p}-B^*({\sf p},\tau)\wbd_{-{\sf p}}\;,
    \end{split}
\end{equation}
where $A$ and $B$ are complex functions to be determined. We require $\wxid_{\sf p}$ 
and $\wxi_{\sf p}$ to fulfill the usual algebra:
\begin{equation}\label{xicommrel}
    [\wxi_{\sf p}(\tau),\wxid_{{\sf p}'}(\tau)]=\delta^2(\pT-\pT')\delta(\mu-\mu')\;,\qquad
    [\wxi_{\sf p}(\tau),\wxi_{{\sf p}'}(\tau)]=0=[\wxid_{\sf p}(\tau),\wxid_{{\sf p}'}(\tau)]\;,
\end{equation}
so that, by enforcing the commutation relations \eqref{commrel}, we find respectively
\begin{equation*}
    \begin{split}
        &\left(|A({\sf p},\tau)|^2-|B({\sf p},\tau)|^2\right)\delta^2(\pT-\pT')\delta(\mu-\mu')=\delta^2(\pT-\pT')\delta(\mu-\mu')\;,\\
        &\left[A^*(-{\sf p},\tau)B^*({\sf p},\tau)-A^*({\sf p},\tau)B^*(-{\sf p},\tau)\right]\delta^2(\pT+\pT')\delta(\mu+\mu')=0\;,\\
        &\left[A({\sf p},\tau)B(-{\sf p},\tau)-A(-{\sf p},\tau)B({\sf p},\tau)\right]\delta^2(\pT+\pT')\delta(\mu+\mu')=0\;.
    \end{split}
\end{equation*}
The above equation is fulfilled if:
\be\label{conditions}
A({\sf p},\tau)=A(-{\sf p},\tau)\;,\qquad B({\sf p},\tau)=B(-{\sf p},\tau)\;,\qquad 
|A({\sf p},\tau)|^2-|B({\sf p},\tau)|^2=1\;,
\ee
so we can set:
\begin{equation}\label{redef}
    A({\sf p},\tau)=\cosh \theta({\sf p},\tau) \,\e^{\ii \chi_{A}({\sf p},\tau)}\;,\qquad
    B({\sf p},\tau)=\sinh \theta({\sf p},\tau) \, \e^{\ii \chi_B({\sf p},\tau)}\;.
\end{equation}
The conditions \eqref{conditions} make it easier to invert Eq.\ \eqref{bogo1}:
\begin{equation}\label{bogo2}
    \begin{split}
        &\wb_{\sf p} = A({\sf p}, \tau)\wxi_{\sf p}(\tau) + B^*({\sf p},\tau)\wxid_{-{\sf p}}(\tau)\;,\\
        &\wbd_{\sf p}= A^*({\sf p},\tau)\wxid_{\sf p}(\tau) + B({\sf p},\tau)\wxi_{-{\sf p}}(\tau)\;.
    \end{split}
\end{equation}
Plugging Eq.\ \eqref{bogo2} into Eq.\ \eqref{Pitau} we obtain
\begin{equation}\label{Pitau2}
    \begin{split}
        \wPi(\tau)=&
        \int{\rm d}^2{\rm p_T}\,{\rm d}\mu\,\frac{\omega}{2}\left\{
        \left[K\left(|A|^2+|B|^2\right)+\Lambda AB^*+\Lambda^*A^*B\right]\left(\wxi_{\sf p}\wxid_{\sf p}+\wxid_{\sf p}\wxi_{\sf p}\right)\right.\\
        &\left.+\left(2KAB+\Lambda A^2+\Lambda^*B^2\right)\wxi_{\sf p}\wxi_{-{\sf p}}+\left(2KA^*B^*+\Lambda^*{A^*}^2+\Lambda{B^*}^2\right)\wxid_{\sf p}\wxid_{-{\sf p}}\right\}\;,
    \end{split}
\end{equation}
where we used the invariance of the integral under reflections ${\sf p}\mapsto -{\sf p}$. In order to
make $\wPi(\tau)$ diagonal, the second line of Eq.\ \eqref{Pitau2} must vanish:
$$
        2 K A B + \Lambda A^2 + \Lambda^* B^2 = 0
$$
(the other equation is just the complex conjugate). This can be rewritten by using Eqs.\ \eqref{reltheta}
and \eqref{redef}:
$$
  \cosh 2 \Theta \sinh 2 \theta \, \e^{\ii (\chi_A + \chi_B)} + \sinh 2\Theta 
  \cosh^2 \theta \, \e^{\ii (\chi + 2 \chi_A)} + \sinh 2\Theta  \sinh^2 \theta \,
  \e^{\ii (2\chi_B -\chi)} = 0 \;,
$$
the solution of which is:
$$
   \chi_B - \chi_A = \chi \;, \qquad \qquad \theta = -\Theta\;.
$$
We can then set $\chi_A=0$ and find $A, B$ fulfilling the Bogolyubov relations \eqref{bogo1}
\be\label{bogosol}
  A = \cosh \Theta \;, \qquad \qquad B = - \sinh \Theta \, \e^{\ii \chi}\;,
\ee
whence, by using Eq.\  \eqref{reltheta}
$$
K\left(|A|^2+|B|^2\right)+\Lambda AB^*+\Lambda^*A^*B = \cosh^2 2 \Theta 
- 2 {\rm Re} \left( \sinh 2 \Theta \, \e^{\ii \chi} \cosh \Theta \sinh \Theta \,
 \e^{-\ii \chi} \right) = 1\;.
$$
With these solutions, Eq.\ \eqref{bogo2} becomes:
\begin{align}\label{bogo3}
 &\wb_{\sf p} = \cosh \Theta({\sf p}, \tau) \wxi_{\sf p}(\tau) - \sinh \Theta ({\sf p},\tau) 
  \e^{-\ii \chi} \wxid_{-{\sf p}}(\tau)\;,\nonumber\\
 &\wbd_{\sf p} = \cosh \Theta({\sf p}, \tau) \wxid_{\sf p}(\tau) - \sinh \Theta ({\sf p},\tau) 
  \e^{\ii \chi} \wxi_{-{\sf p}}(\tau) \;,
\end{align}
and the operator \eqref{Pitau2}:
\begin{equation}\label{Pitau3}
    \wPi(\tau)=\int{\rm d}^2{\rm p_T}\,{\rm d}\mu\,\frac{\omega}{2}
    \left(\wxi_{\sf p}(\tau)\wxid_{\sf p}(\tau)+\wxid_{\sf p}(\tau) \wxi_{\sf p}(\tau)
     \right) = \int{\rm d}^2{\rm p_T}\,{\rm d}\mu\,\omega 
      \left(\wxid_{\sf p}(\tau) \wxi_{\sf p}(\tau) + \frac{1}{2} \right)\;,
\end{equation}
where in the last equality we have used the commutation relations \eqref{xicommrel}.

\subsection{Discusssion}

The non-trivial Bogoliubov transformation \eqref{bogo3} between different sets of creation and annihilation 
operators is reminiscent of the Unruh effect \cite{crispino} and indeed the velocity field $u$ implied in
Eq.\ \eqref{beta} has non-vanishing acceleration. However, we are facing essentially different physics here; 
as it has been pointed out, the relation \eqref{atob}  between plane-wave creation operators and the creation operators 
appearing in the field expansion in curvilinear coordinates does not mix creation and
annihilation operators. In other words, unlike in the Unruh effect, the observers associated with Milne 
coordinates (defined by $\eta={\bf x}_{\rm T}=const$) count the same particles as the inertial observer.

In fact, the Bogolyubov transformation \eqref{bogo3} stems from the somewhat unexpected form of the local 
thermodynamic equilibrium operator $\wPi$ in Eq.\ \eqref{Pitau} involving quadratic combinations of 
two annihilation and two creation operators, unlike the Hamiltonian in global-equilibrium thermal field 
theory. We thus have a concrete situation where the vacuum $\ket{0_\tau}$, which is the lowest-lying eigenvector 
of $\wPi(\tau)$ annihilated by all $\wxi_{{\sf p}}(\tau)$'s,
$$
  \wxi_{{\sf p}}(\tau) \ket{0_\tau} = 0\;,
$$
is different from the Minkowski vacuum $\ket{0_M}$, which is annihilated by the $\wb_{{\sf p}}$, as envisioned 
in Sec.~\ref{vacuum}. The full expression of the vacuum $\ket{0_\tau}$ can be obtained from the 
coefficients in Eq.\ \eqref{bogo3} with known methods \cite{winitzki} and reads:
\be\label{twovacua}
\ket{0_\tau} = \prod_{{\sf p}} \frac{1}{|\cosh \Theta({\sf p},\tau)|^{1/2}} 
 \exp \left[ -\frac{1}{2}\tanh\Theta({\sf p},\tau) \e^{-\ii \chi({\sf p},\tau)} \wbd_{\sf p} \wbd_{\sf -p} 
 \right] \ket{0_M}\;.
\ee

With $\wPi$ diagonal in Eq.~\eqref{Pitau3}, we can readily obtain the expectation values of products of creation and 
annihilation operators in local thermodynamic equilibrium. The form \eqref{Pitau3} is essentially the same as the 
equilibrium Hamiltonian operator of the free field with the replacements $\mu \to p_z$ and $\omega \to \varepsilon$. 
We thus have:
\begin{equation}\label{ximean}
    \begin{split} 
 \langle \wxid_{\sf p}(\tau)\wxi_{{\sf p}'}(\tau)\rangle_{\rm LE}=&n_{\rm B}({\sf p},\tau)\,
 \delta^2(\pT-\pT')\delta(\mu-\mu')\;, \\
        \langle \wxi_{\sf p}(\tau)\wxid_{{\sf p}'}(\tau)\rangle_{\rm LE}=&
        \left[n_{\rm B}({\sf p},\tau)+1\right]\delta^2(\pT-\pT')\delta(\mu-\mu')\;,\\
        \langle \wxi_{\sf p}(\tau)\wxi_{{\sf p}'}(\tau)\rangle_{\rm LE}=&
        0=\langle \wxid_{\sf p}(\tau)\wxid_{{\sf p}'}(\tau)\rangle_{\rm LE}\;,
    \end{split}
\end{equation}
where $\langle \cdot \rangle_{\rm LE}$ stands for $\Tr (\wrhol \cdot )$ and $n_{\rm B}$ is the Bose-Einstein 
distribution function:
\begin{equation}\label{bose}
  n_{\rm B}({\sf p},\tau)=\frac{1}{{\rm e}^{\omega(\tau)/T(\tau)}-1}\;,
\end{equation}
with $\omega(\tau)$ given by Eq.\ \eqref{pitau}. 

It is important to emphasize that Eq.\ \eqref{bose} is by no means a density of particles as usually 
in Minkowski space-time. Equation \eqref{bose} accounts for the mean number of excitations of 
the $\wxid_{\sf p}(\tau)$ operator, which is {\em not} the mean number of excitations of the Minkowski vacuum as 
expressed by the $\wa{p}$'s or $\wb_{\sf p}$'s. Indeed, the expectation values of the various combinations 
can be found by means of Eq.\ \eqref{bogo2} including the solution \eqref{bogosol} and Eq.\ \eqref{ximean}:
\begin{align}\label{bbmean}
 \media{\wb_{\sf p}\wb_{\sf p^\prime}}_{\rm LE} &= - \frac{1}{2} \sinh (2\Theta) \, \e^{-\ii \chi} (2 n_{\rm B} +1)
  \delta^2(\pT-\pT')\delta(\mu-\mu')\;, \\ \nonumber
  \media{\wbd_{\sf p}\wbd_{\sf p^\prime}}_{\rm LE} &= - \frac{1}{2} \sinh (2\Theta) \, \e^{\ii \chi} (2 n_{\rm B} +1)
  \delta^2(\pT-\pT')\delta(\mu-\mu')\;, \\ \nonumber
 \media{\wb_{\sf p}\wbd_{\sf p^\prime}}_{\rm LE} &= \left[ n_{\rm B} \cosh (2\Theta)  + \cosh^2 \Theta 
  \right] \delta^2(\pT-\pT')\delta(\mu-\mu')\;, \\ \nonumber
 \media{\wbd_{\sf p}\wb_{\sf p^\prime}}_{\rm LE} &= \left[ \, n_{\rm B} \cosh (2\Theta)  + \sinh^2 \Theta 
  \right] \delta^2(\pT-\pT')\delta(\mu-\mu')  \;.
\end{align}
As is clear from Eq.\ \eqref{bbmean}, field vacuum effects are encoded in a non-vanishing value of
the angle $\Theta({\sf p},\tau)$, which is both a function of the modes and of the Milne time $\tau$ and
whose value can be determined through the relations \eqref{reltheta}. We are now in a position to 
calculate the expectation values of all operators which are quadratic in the field.

\section{The stress-energy tensor and its renormalization}
\label{setensor}

We now come to the main point of this work, namely the determination of the stress-energy tensor. We start
by calculating it in local thermodynamic equilibrium. 

\subsection{Local thermodynamic equilibrium}

As the symmetries of $\wrhol$ are the same as $\wrho$ 
(see the discussion in Sec.\  \ref{symmetry}) the structure must be the same as in Eq.\  \eqref{set}:
\begin{equation}\label{setleq}
   \Tr(\wrhol \wT^{\mu\nu}) = \langle \wT^{\mu \nu} \rangle_{\rm LE} 
   = {\cal E}(\tau)_{\rm LE} u^{\mu}u^{\nu}+{\cal P}_{\rm T}(\tau)_{\rm LE}\left(\hat i^{\mu} \hat i^{\nu}+ \hat j^{\mu} \hat j^{\nu}\right)+{\cal P}_{\rm L}(\tau)_{\rm LE} \hat \eta^{\mu} 
     \hat \eta^{\nu}\;,
\end{equation}
Hence, by using Eq.\  \eqref{setmilne} with the expansion \eqref{field} we obtain:
\begin{equation}\label{rhole}
    \begin{split}
        {\cal E}(\tau)_{\rm LE}=\langle \wT_C^{\mu \nu}\rangle_{\rm LE}u_{\mu}u_{\nu}=&
        \int\frac{{\rm d^2p_T}\,{\rm d}\mu\,{\rm d^2p_T'}\,{\rm d}\mu'}{4(4\pi)^2} \\
        &\times \left(\left\{ [\partial_{\tau}h({\sf p},\tau)][\partial_{\tau}h({\sf p}',\tau)]-\left(p_xp_x'+p_yp_y'+\frac{1}{\tau^2}\mu\mu'-m^2\right)h({\sf p},\tau)h({\sf p}',\tau)\right\} \right.\\
        &\hspace*{0.5cm}\times {\rm e}^{i[(\pT+\pT')\cdot {\bf x}_{\rm T}+(\mu+\mu')\eta]}\langle \wb_{\sf p}\wb_{{\sf p}'}\rangle_{\rm LE}\\
        &\hspace*{0.25cm}+\left\{[\partial_{\tau}h({\sf p},\tau)][\partial_{\tau}h^*({\sf p}',\tau)]+\left(p_xp_x'+p_yp_y'+\frac{1}{\tau^2}\mu\mu'+m^2\right)h({\sf p},\tau)h^*({\sf p}',\tau)\right\} \\
        &\hspace*{0.5cm}\times {\rm e}^{i[(\pT-\pT')\cdot {\bf x}_{\rm T}+(\mu-\mu')\eta]}\langle \wb_{\sf p}\wb_{{\sf p}'}^{\dagger}\rangle_{\rm LE}\\
        &\hspace*{0.25cm}+\left\{[\partial_{\tau}h^*({\sf p},\tau)][\partial_{\tau}h({\sf p}',\tau)]+\left(p_xp_x'+p_yp_y'+\frac{1}{\tau^2}\mu\mu'+m^2\right)h^*({\sf p},\tau)h({\sf p}',\tau)\right\} \\
        &\hspace*{0.5cm}\times {\rm e}^{-i[(\pT-\pT')\cdot {\bf x}_{\rm T}+(\mu-\mu')\eta]}\langle \wb_{\sf p}^{\dagger}\wb_{{\sf p}'}\rangle_{\rm LE}\\
        &\hspace*{0.25cm}+\left\{[\partial_{\tau}h^*({\sf p},\tau)][\partial_{\tau}h^*({\sf p}',\tau)]-\left(p_xp_x'+p_yp_y'+\frac{1}{\tau^2}\mu\mu'-m^2\right)h^*({\sf p},\tau)h^*({\sf p}',\tau)\right\} \\
        &\hspace*{0.5cm}\left.\times {\rm e}^{-i[(\pT+\pT')\cdot {\bf x}_{\rm T}+(\mu+\mu')\eta]}\langle \wb_{\sf p}^{\dagger}\wb_{{\sf p}'}^{\dagger}\rangle_{\rm LE}\right)\;.
    \end{split}
\end{equation}
Plugging the relations \eqref{bbmean} into Eq.~\eqref{rhole} we obtain:
\begin{eqnarray*}
 {\cal E}(\tau)_{\rm LE}& =&  \frac{1}{4(4\pi)^2} \int {\rm d^2 p_T}\,{\rm d}\mu
 \left\{ -\frac{1}{2} \left[ (\partial_{\tau}h)^2+\omega^2 h^2 \right]
  \left( 2n_{\rm B}+1 \right) \sinh (2 \Theta) \, \e^{-\ii \chi} + {\rm c.c.} \right. \\
&  & \hspace*{3.1cm}  + \left. \left( |\partial_{\tau}h|^2+\omega^2 |h|^2 \right) \left( 2 n_{\rm B} + 1 \right) 
  \cosh (2 \Theta) \frac{}{} \right\}\;,
\end{eqnarray*}
and, by using Eqs.~\eqref{kappa}, \eqref{lambda}, and \eqref{reltheta}:
\be\label{rhofinal}
    {\cal E}(\tau)_{\rm LE} = \frac{1}{16\pi^3 \tau} \int {\rm d^2p_T}\,{\rm d}\mu \; 
    \omega \left[(2 n_{\rm B} + 1) \cosh^2 (2 \Theta) - \sinh^2 (2 \Theta) (2 n_{\rm B} + 1) \right] 
    = \frac{1}{(2\pi)^3 \tau} \int \di^2 {\rm p_T} \di \mu \; \omega \left( n_{\rm B} + 
    \frac{1}{2} \right)\;.
\ee
The longitudinal and transverse pressures can be worked out in a similar fashion: the Wronskian of 
the Hankel function is again recovered and the expressions greatly simplify. One obtains:
\begin{equation}\label{pressures}
    \begin{split}
        {\cal P}_{\rm T}(\tau)_{\rm LE}=& \frac{1}{(2\pi)^3 \tau} \int \di^2 {\rm p_T} \di \mu \; \frac{|\pT|^2}{2}\left(n_{\rm B}+\frac{1}{2}\right)\;, \\
        {\cal P}_{\rm L}(\tau)_{\rm LE}=&\frac{1}{(2\pi)^3 \tau} \int \di^2 {\rm p_T} \di \mu \; \frac{\mu^2}{\tau^2}\left(n_{\rm B}+\frac{1}{2}\right)\;. 
    \end{split}
\end{equation}
Equations \eqref{rhofinal} and  \eqref{pressures} can be written in a compact fashion by introducing
the functions:
\begin{equation}\label{kappa2}
    K_\gamma({\sf p},\tau)=\frac{\pi \tau}{4 \omega} \left[|\partial_{\tau}h({\sf p},\tau)|^2
    +\gamma({\sf p},\tau)|h({\sf p},\tau)|^2\right]\;,
\end{equation}
\begin{equation}\label{lambda2}
    \Lambda_\gamma({\sf p},\tau)=\frac{\pi \tau}{4 \omega}
    \left\{ [\partial_{\tau}h({\sf p},\tau)]^2+\gamma({\sf p},\tau)[h({\sf p},\tau)]^2\right\}\;,
\end{equation}
where $\gamma$ is defined as
\be\label{gammas}
    \gamma({\sf p},\tau)=\left\{
    \begin{split}
        \omega^2({\sf p},\tau)= m_{\rm T}^2 + \frac{\mu^2}{\tau^2}\;,\qquad \qquad &\mbox{for}\quad {\cal E}(\tau)_{\rm LE}\;,\\
        -m_{\rm L}^2 \equiv -\frac{\mu^2}{\tau^2}-m^2\;,\qquad \qquad &\mbox{for}\quad {\cal P}_{\rm T}(\tau)_{\rm LE}\;,\\
        -m_{\rm T}^2 + \frac{\mu^2}{\tau^2}\;,\qquad \qquad &\mbox{for}\quad {\cal P}_{\rm L}(\tau)_{\rm LE}\; .
    \end{split}
    \right.
\ee
Thanks to the Wronskian of the Hankel functions, they satisfy the relation:
$$
    K_\gamma^2({\sf p},\tau)-|\Lambda_\gamma({\sf p},\tau)|^2=\frac{\gamma({\sf p},\tau)}
    {\omega^2({\sf p},\tau)}\;.
$$
With this in mind, and setting $\Gamma_\gamma = \{ {\cal E},{\cal P}_{\rm T},{\cal P}_{\rm L}\}$, we have for
the thermodynamic function of the stress-energy tensor:
$$
    \Gamma_\gamma(\tau)_{\rm LE} = \int \frac{{\rm d^2p_T}\,{\rm d}\mu}{(2\pi)^3\tau}\omega({\sf p},\tau)
    \left[K_\gamma({\sf p},\tau) K({\sf p},\tau)-{\rm Re} \left(\Lambda_\gamma({\sf p},\tau)
    \Lambda^*({\sf p},\tau)\right)\right]\left(n_{\rm B}({\sf p},\tau)+\frac{1}{2}\right)\;,
$$
where the combination in square brackets reads
$$
    K_{\gamma}({\sf p},\tau)K({\sf p},\tau)-{\rm Re}\left(\Lambda{\gamma}({\sf p},\tau)\Lambda^*({\sf p},\tau)\right)=\frac{\omega^2({\sf p},\tau)+\gamma({\sf p},\tau)}{2\omega^2({\sf p},\tau)}\;,
$$
hence
\begin{equation}\label{gammale}
    \Gamma_{\gamma}(\tau)_{\rm LE}=\int \frac{{\rm d^2p_T}\,{\rm d}\mu}{(2\pi)^3\tau \omega({\sf p},\tau)}\frac{\omega^2({\sf p},\tau)+\gamma({\sf p},\tau)}{2}\left[n_{\rm B}({\sf p},\tau)+\frac{1}{2}\right]\;.
\end{equation}
The above integrals can be written in a familiar form by changing the integration variable to $p_z=\mu/\tau$.
This implies:
$$
 \omega({\sf p},\tau)=\sqrt{|\pT|^2+\frac{\mu^2}{\tau^2}+m^2}=\sqrt{\p_x^2+\p_y^2+\p_z^2+m^2} = \varepsilon\;,
$$
which is just the on-shell energy, and: 
$$
  {\rm d}^2{\rm p_T}\,\frac{{\rm d}\mu}{\tau} = \di \p_x\, \di \p_y\, \di \p_z \;,
$$
In turn, the distribution $n_{\rm B}({\sf p},\tau)$ becomes the energy-dependent Bose-Einstein phase-space distribution $n_{\rm B}(\varepsilon,T(\tau))$. Hence, the first term of the energy density \eqref{rhofinal} as 
well as the transverse and longitudinal pressures \eqref{pressures} can be written as the familiar
momentum integrals of the relativistic uncharged Bose gas. Altogether, the unrenormalized stress-energy
tensor in local equilibrium reads:
\be\label{tmunule}
  \Tr \left[\wrhol \wT^{\mu\nu}(x)\right] = \int \frac{\di^3 p}{\varepsilon} p^\mu p^\nu \left[ 
   \frac{1}{\e^{\beta(x)\cdot p}-1} + \frac{1}{2} \right]\;,
\ee
where $\beta$ is the four-temperature in Eq.\  \eqref{beta}. Hence, the thermodynamic functions 
$\Gamma_\gamma$ are just the familiar functions of $T(\tau)$ as for the ideal 
relativistic gas. In particular, the transverse and the longitudinal pressures are in fact identical, namely
\be\label{pressureiso}
    {\cal P}_{\rm T}(\tau)_{\rm LE}={\cal P}_{\rm L}(\tau)_{\rm LE} \equiv {\cal P}(\tau)_{\rm LE}\; .
\ee
%

\subsection{Actual stress-energy tensor}

The actual (unrenormalized) expectation value of the stress-energy tensor can be calculated by using the density 
operator \eqref{nedo3}, that is:
$$
  \Tr \left[\wrho \, \wT^{\mu\nu}(x)\right] = \frac{1}{Z} \Tr \left\{\exp[-\wPi(\tau_0)/T(\tau_0)]\wT^{\mu\nu}(x) \right\}\;.
$$
Symmetries dictate that its form is given by Eq.~\eqref{set}, so we need to determine the
three functions $\Gamma_\gamma$. It is readily found that the same expression as in Eq.\ \eqref{rhole} is obtained, with the simple replacement of the local-equilibrium values of the quadratic combinations
of $\wb_{\sf p}$ and $\wbd_{\sf p}$  with their actual expectation values, for instance:
$$
  \media{\wbd_{\sf p} \wb_{\sf p}} =  \frac{1}{Z} \Tr \left\{ \exp[-\wPi(\tau_0)/T(\tau_0)]\,\wbd_{\sf p}
   \wb_{\sf p}\right\}\;.
$$
The calculation of the above expression is most easily done by using the formulae \eqref{bogo3} {\em at 
time $\tau_0$}, i.e., expressing the constant $\wb_{\bf p}$'s as  functions of the operators diagonalizing
$\wPi(\tau_0)$ instead of $\wPi(\tau)$. We thus get the same formulae as Eq.\ \eqref{bbmean}, with 
$\tau$ replaced by $\tau_0$:
\begin{align}\label{bbmeantrue}
 \media{\wb_{\sf p}\wb_{\sf p^\prime}} &= - \frac{1}{2} \sinh [2\Theta(\tau_0)] \, 
 \e^{-\ii \chi (\tau_0)} [2 n_{\rm B}(\tau_0) +1]
  \delta^2(\pT-\pT')\delta(\mu-\mu') \;, \\ \nonumber
  \media{\wbd_{\sf p}\wbd_{\sf p^\prime}} &= - \frac{1}{2} \sinh [2\Theta(\tau_0)] \, 
  \e^{\ii \chi(\tau_0)} [2 n_{\rm B}(\tau_0) +1]
  \delta^2(\pT-\pT')\delta(\mu-\mu')\;, \\ \nonumber
 \media{\wb_{\sf p}\wbd_{\sf p^\prime}} &= \left\{ n_{\rm B}(\tau_0) \cosh [2\Theta (\tau_0)] + 
  \cosh^2 \Theta (\tau_0) \right\} \delta^2(\pT-\pT')\delta(\mu-\mu')\;, \\ \nonumber
 \media{\wbd_{\sf p}\wb_{\sf p^\prime}} &= \left\{  n_{\rm B}(\tau_0) \cosh [2\Theta(\tau_0)] + 
 \sinh^2 \Theta (\tau_0) \right\} \delta^2(\pT-\pT')\delta(\mu-\mu')  \;.
\end{align}
We note in passing that, as expected, the expectation value of excitations of the Minkowski vacuum, described by 
$\media{\wbd_{\sf p}\wb_{\sf p}}$ for each mode, is constant in time, the density operator being
fixed and the operators $\wb_{\sf p}$ being time-independent by construction. The mean number of particles
with momentum $p$ can be obtained by using Eq.\ \eqref{atob}:
$$
 \media{\wad{p} \wa{p^\prime}} = \frac{1}{2\pi m_{\rm T}} \frac{1}{\sqrt{\cosh y \cosh y'}} 
 \int_{-\infty}^{+\infty} \di \mu \; \e^{-\ii \mu (y-y^\prime)} \media{\wbd_{{{\bf p}_T},\mu}
 \wb_{{{\bf p}_T^\prime},\mu}}\;,
$$
where $y$ is the rapidity. 

Now, by taking advantage of the right-hand side of Eq.\ \eqref{rhole}, and by using Eq.\  \eqref{bbmeantrue}, 
as well as Eqs.\  \eqref{kappa}, \eqref{lambda}, and \eqref{reltheta}, it can be shown that:
\begin{equation}
    {\cal E}(\tau)=  \frac{1}{Z} \Tr \left\{\exp[-\wPi(\tau_0)/T(\tau_0)]\wT^{\mu\nu}\right\} u_\mu u_\nu = 
    \frac{1}{(2 \pi)^3 \tau} \int {\rm d^2 p_T} \,{\rm d}\mu \; \omega(\tau)
    \left\{ K(\tau)K(\tau_0)- {\rm Re} \left[ \Lambda(\tau)\Lambda^*(\tau_0) \right]\right\}
    \left[n_{\rm B}(\tau_0)+\frac{1}{2}\right]\;.
\end{equation}
The pressures can be derived likewise and we finally have:
\begin{equation}\label{gammatrue}
    \Gamma_\gamma(\tau) = \frac{1}{(2\pi)^3\tau}\int {\rm d^2p_T} \,{\rm d}\mu \; \omega(\tau)
    \left\{K_\gamma(\tau) K(\tau_0) - {\rm Re} \left[ \Lambda_\gamma(\tau)\Lambda^*(\tau_0)\right]
    \right\}\left[n_{\rm B}(\tau_0)+\frac{1}{2}\right]\;.
\end{equation}
Of course, at the time $\tau=\tau_0$ we recover the local thermodynamic equilibrium expression \eqref{gammale}, 
as required by construction. However, at later times $\tau > \tau_0$ the stress-energy tensor 
differs from the local equilibrium form. Indeed, since we are dealing with a free field, one expects to find
the same expression as for the free-streaming solution of the Boltzmann equation in Milne coordinates, see
Appendix \ref{freestreaming}. However, there are quantum corrections due to the vacuum subtraction.

\subsection{Renormalization and comparison with classical limits}

The expressions found include divergent terms, both in the stress-energy tensor in local equilibrium
\eqref{gammale} and the actual one \eqref{gammatrue}. As we have seen in Sec.\ \ref{vacuum},
in order to fulfill the continuity equation, the stress-energy tensor should be renormalized by subtracting a 
vacuum expectation value (VEV) with a constant vacuum: either with respect to the Minkowskian vacuum $\ket{0_M}$, like in Eq.~\eqref{minksub}, or with respect to the vacuum $\ket{0_{\tau_0}}$ of the operator $\wPi(\tau_0)$, like in Eq.~\eqref{tausub}. 

The Minkowski VEV of the stress-energy tensor is calculated in Appendix \ref{minkoswkivev}. For the 
stress-energy tensor it is found:
$$
   {\cal E}_M  \equiv \bra{0_M} \wT^{\mu\nu} u_\mu u_\nu \ket{0_M} =
   \frac{1}{(2 \pi)^3 \tau} \int {\rm d^2 p_T} \,{\rm d}\mu \; \omega(\tau)\, \frac{K(\tau)}{2}\;,
$$
and the renormalized energy density is then:
$$
    {\cal E}(\tau)_{\rm ren} = \frac{1}{(2 \pi)^3 \tau} \int {\rm d^2 p_T} \,{\rm d}\mu \; \omega(\tau)
    \left( \left\{ K(\tau)K(\tau_0)- {\rm Re} \left[ \Lambda(\tau)\Lambda^*(\tau_0) \right]\right\}
    \left[ n_{\rm B}(\tau_0) + \frac{1}{2} \right] - \frac{1}{2} K(\tau) \right)\;.  
$$
The main drawback of this expression is that it is still divergent. This is most easily seen at $\tau=\tau_0$
where:
\be\label{evacmin}
  {\cal E}(\tau_0)_{\rm ren} = \frac{1}{(2 \pi)^3 \tau_0} \int {\rm d^2 p_T} \,{\rm d}\mu \; \omega(\tau_0)
   n_{\rm B}(\tau_0) - \frac{1}{2(2 \pi)^3 \tau_0} \int {\rm d^2 p_T} \,{\rm d}\mu \; \omega(\tau_0) \left[ 
   K(\tau_0) - 1 \right]\;.
\ee
While the first term is finite, the second is not due to the behaviour of the $K$ function for large
values of its effective argument, which is $m_T \tau_0$, at fixed $\mu$ [see Eqs.~\eqref{kappa} and \eqref{hankel}].
The asymptotic behaviour for large transverse mass $m_{\rm T}$  of the $K$ function is derived
in Appendix \ref{asymptotics} and one has, at leading order:
$$
  K(\tau_0) - 1 = \cosh \Theta(\tau_0) -1 \simeq \frac{\Theta^2}{2} \simeq 
  \frac{1}{8m_{\rm T}^2\tau_0^2}\;,
$$
which makes the integral in Eq.\  \eqref{evacmin} divergent. 

In conclusion, in order to have a finite stress-energy tensor, we are left with the option to subtract 
the VEV's with respect to $\ket{0_{\tau_0}}$, which can be readily done by taking the limit $T(\tau_0)\to 0$ in Eq.\ 
\eqref{gammatrue} and subtracting what is left, taking into account that $ \lim_{T \to 0} n_B = 0$. We 
thus have:
\begin{equation}\label{gammaren}
    \Gamma_\gamma(\tau)_{\rm ren} = \frac{1}{(2\pi)^3\tau}\int {\rm d^2p_T} \,{\rm d}\mu \; \omega(\tau)
    \left\{K_\gamma(\tau) K(\tau_0) - {\rm Re} \left[ \Lambda_\gamma(\tau)\Lambda^*(\tau_0)\right]
    \right\} n_{\rm B}(\tau_0)\;,
\end{equation}
which incorporates the relation between the energy density and the pressures.

It is interesting to study the behaviour of the functions \eqref{gammaren} at late times $\tau$, which means
for large values of $m \tau$  (see Appendix \ref{asymptotics}). In this limit, we have $\Theta(\tau) \to 0$, 
hence $K(\tau) \to 1$ and $\Lambda(\tau) \to 0$, implying that the Minkowskian vacuum is recovered asymptotically. 
This is also clear from Eq.~\eqref{twovacua}, which shows that $\ket{0_\tau} \to \ket{0_M}$. For the 
energy density, at late times we have:
\begin{align}\label{rholate}
    {\cal E}(\tau)_{\rm ren} &\underset{\tau \to \infty}{\simeq} 
     \frac{1}{(2\pi)^3\tau}\int {\rm d^2p_T} \,{\rm d}\mu \; \omega(\tau) K(\tau_0) n_{\rm B}(\tau_0)  \\ \nonumber
     & = \frac{1}{(2\pi)^3\tau}\int {\rm d^2p_T} \,{\rm d}\mu \; 
     \omega(\tau) n_{\rm B}(\tau_0) + \frac{1}{(2\pi)^3\tau}\int {\rm d^2p_T} \,{\rm d}\mu \; 
     \omega(\tau) [K(\tau_0) - 1] n_{\rm B}(\tau_0) \\ \nonumber
   &= \frac{1}{(2\pi)^3\tau}\int {\rm d^2p_T} \,{\rm d}\mu \; 
     \omega(\tau) n_{\rm B}(\tau_0) + \frac{2}{(2\pi)^3\tau}\int {\rm d^2p_T} \,{\rm d}\mu \; 
     \omega(\tau) \sinh^2 \Theta(\tau_0) n_{\rm B}(\tau_0)\;.
\end{align}
It can be shown that the first term in Eq.~\eqref{rholate} is the classical free-streaming solution 
in Milne coordinates (see Appendix \ref{freestreaming}), while the second term is a pure quantum-field correction 
due to the difference between vacua, since it vanishes only if $\Theta(\tau_0)=0$. Somewhat surprisingly, the 
quantum correction to energy density does not vanish at late times, and it can even be comparable with 
the classical term if the main argument of $\Theta(\tau_0)$, that is $m_T \tau_0$ is ${\cal O}(1)$, that 
is for an early decoupling of the system. 

Similar expressions can be obtained for the pressures. For large times, the leading term of the
$\Lambda_\gamma(\tau)$ function has an oscillating behaviour $\sim \exp(-2 \ii m_T \tau)$ [see Appendix 
\ref{asymptotics}, Eq.~\eqref{lambda3}], so the integrals in ${\rm p_T}$ or $m_{\rm T}$ involving $\Lambda_\gamma(\tau)$ 
are expected to decay as $\tau \to \infty$. Therefore, only the first term of Eq.\ \eqref{gammaren} is left 
and one has:
\be\label{presslate}
    {\cal P}_\gamma (\tau)_{\rm ren} \underset{\tau \to \infty}{\simeq} 
     \frac{1}{(2\pi)^3\tau}\int {\rm d^2 p_T} \,{\rm d}\mu \; \omega(\tau) K_\gamma(\tau) 
     K (\tau_0) n_{\rm B}(\tau_0) \;.
\ee
Also, at late times [see Appendix \ref{asymptotics}, Eq.~\eqref{kappa3}]:
$$
  \omega(\tau) K_\gamma (\tau) \underset{\tau \to \infty}{\simeq} \frac{m^2_{\rm T} + \gamma}{2 m_{\rm T}}\;,
$$  
so Eq.\ \eqref{presslate} becomes:
\begin{align}\label{presslate2}
   {\cal P}_\gamma (\tau)_{\rm ren} &\underset{\tau \to \infty}{\simeq} 
    \frac{1}{(2\pi)^3\tau}\int {\rm d^2p_T} \,{\rm d}\mu \; \frac{m^2_{\rm T} + \gamma}{2 m_{\rm T}}
     K (\tau_0) n_{\rm B}(\tau_0) \\ \nonumber
   &= \frac{1}{(2\pi)^3\tau}\int {\rm d^2p_T} \,{\rm d}\mu \; 
    \frac{m^2_{\rm T} + \gamma}{2 m_{\rm T}} n_{\rm B}(\tau_0) + 
    \frac{2}{(2\pi)^3\tau}\int {\rm d^2p_T} \,{\rm d}\mu \; 
    \frac{m^2_{\rm T} + \gamma}{2 m_{\rm T}} \sinh^2 \Theta(\tau_0) n_{\rm B}(\tau_0)\;,
\end{align}
with  $\gamma$ from Eq.~\eqref{gammas}.
Again, the first term is the leading approximation of the classical free-streaming solution in 
Milne coordinates for large $m_T \tau$ and fixed $\mu$, whereas the second term is a pure quantum 
correction. 

\section{Entropy current}
\label{entropycurrent}

The need of subtracting the vacuum $\ket{0_{\tau_0}}$ to obtain a finite value for the stress-energy
tensor for the free field has some interesting connection to the way the entropy and the entropy current
of a relativistic fluid in local thermodynamic equilibrium are calculated. This problem has been approached 
in the framework of the relativistic density operator in Ref.~\cite{becarindo}. We first observe
that the entropy of a relativistic fluid in local equilibrium,
$$
    S=-\tr (\wrho_{\rm LE}\log \wrho_{\rm LE})\;,
$$
with $\wrhol$ given by the Eq.\ \eqref{rhole}, is independent of the vacuum subtraction because,
as remarked in Sec.~\ref{symmetry}, the density operator \eqref{rhole} turns out to be independent 
of any non-operator term which is subtracted from the stress-energy tensor operator, as it cancels
out in the ratio with the normalizing $Z_{\rm LE}$.

However, it was pointed out in Ref.~\cite{becarindo} that, provided that the vacuum is non-degenerate, there 
is only one good choice of the vacuum if one has to make $\log Z_{\rm LE}$ extensive, i.e.:
$$
  \log Z_{\rm LE} = \int_\Sigma \di \Sigma_\mu \phi^\mu\;,
$$
and this is the vacuum (meant as the eigenvector with minimal eigenvalue) of the operator $\wPi(\tau)$,
which we have denoted with $\ket{0_\tau}$. Therefore, \footnote{In this section, for the sake of simplicity, 
we assume vanishing chemical potentials, that is $\zeta=0$; the extension of these arguments to 
a non-vanishing chemical potential is straightforward.} the entropy current reads:
\be\label{entcurr}
  s^\mu = \phi^\mu + \left[ \Tr (\wrhol \wT^{\mu\nu}) - \bra{0_\tau}\wT^{\mu\nu} \ket{0_\tau} \right] 
   \beta_\nu \;,
\ee
with:
$$
    \phi^\mu = \int_1^\infty \di \lambda \; 
    \left\{ \Tr [\wrhol(\lambda) \wT^{\mu\nu}] - \bra{0_\tau}\wT^{\mu\nu} \ket{0_\tau} \right\}\;,
$$    
where $\wrhol(\lambda)$ is the operator defined by:
$$
 \wrho_{\rm LE}(\lambda) = \frac{1}{Z_ {\rm LE}(\lambda)} \exp\left( -\lambda 
 \int_\Sigma \di \Sigma_\mu \; \wT^{\mu\nu} \beta_\nu \right)\;.
$$

The renormalized value: 
$$
 T^{\mu\nu}_{\rm LE} = \Tr (\wrhol \wT^{\mu\nu}) - \bra{0_\tau}\wT^{\mu\nu} \ket{0_\tau}
$$
of the stress-energy tensor in local thermodynamic equilibrium with subtraction of the VEV with respect to $\ket{0_\tau}$ 
can be found by taking the limit $T(\tau) \to 0$, as we have seen in Sec.~\ref{vacuum}. Hence, for the 
free scalar field, it is readily found from Eq.\ \eqref{tmunule} that we are left with the classical expression:
$$
  T^{\mu\nu}(x)_{\rm LE} = \int \frac{\di^3 p}{\varepsilon} p^\mu p^\nu 
   \frac{1}{\e^{\beta(x)\cdot p}-1}\;. 
$$
It is now easy to show that $\phi^\mu = {\cal P}_{\rm LE} \beta^\mu$, with ${\cal P}_{\rm LE}$ being the 
one pressure in Eq.~\eqref{pressureiso}, and that the entropy current coincides with the classical 
equilibrium expression:
$$
 s^\mu = ({\cal E}_{\rm LE} + {\cal P}_{\rm LE}) \beta^\mu\;,
$$
where ${\cal E}_{\rm LE}$ and ${\cal P}_{\rm LE}$ are related by the usual equation of state of a 
free relativistic gas, without apparent quantum correction.

We end this section by discussing the entropy-production rate equation established in Refs.~\cite{weert,zubarev} 
[for a derivation see Ref.~\cite{becazuba}], which for $\zeta=0$ reads:
\begin{equation}\label{prodrate}
    \nabla_{\mu}s^{\mu}=\left(T^{\mu \nu}-T^{\mu \nu}_{\rm LE} \right)\nabla_{\mu}\beta_{\nu}\;.
\end{equation}
In the above equation it is usually understood that $T^{\mu\nu}$ and $T^{\mu \nu}_{\rm LE}$ are the 
renormalized stress-energy tensor expectation values, fulfilling the constraint equation \eqref{constr}, 
and usually obtained by subtracting the Minkowski VEV of both. However, in our case, in order to obtain 
finite values for the constraint equation \eqref{constr} and to find an appropriate expression 
of the entropy current, we need to subtract different VEV's, as we have seen. In particular:
\begin{align*}
    T^{\mu\nu}_{\rm LE} &= \Tr (\wrhol \wT^{\mu\nu}) -  \bra{0_\tau}\wT^{\mu\nu} \ket{0_\tau} \;,\\
    T^{\mu\nu} &= \Tr (\wrho \wT^{\mu\nu}) -  \bra{0_{\tau_0}}\wT^{\mu\nu} \ket{0_{\tau_0}}\;.
\end{align*}
One may thus wonder whether such a difference in the VEV subtraction introduces a new quantum term
in the entropy production rate. The answer is again no, provided that
\begin{itemize}
    \item{} the renormalized expectation value $T^{\mu\nu}$ is finite;
    \item{} the renormalized expectation value $T^{\mu\nu}$ fulfills the continuity equation;
    \item{} the renormalized expectation value in local equilibrium $T^{\mu\nu}_{\rm LE}$ fulfills 
    the constraint \eqref{constr}.
\end{itemize}
The proof of Eq.\ \eqref{prodrate} \cite{becazuba} can be shown to hold.

\section{Summary and conclusions}
\label{conclusions}

To summarize, we have studied a relativistic quantum fluid with longitudinal boost invariance, which, 
for the free scalar field, is an exactly solvable non-equilibrium problem, further developing and extending the results of Refs.~\cite{Akkelin1,Akkelin2}. By using the non-equilibrium density operator, we have derived 
an exact solution for the stress-energy tensor and the entropy current for the free scalar field initially
in local thermodynamic equilibrium. The most remarkable feature of the solution is the difference 
between the vacuum of the density operator and the familiar vacuum of the field in Minkowski space-time. 
We have found that a finite, renormalized value of the stress-energy tensor can be achieved only by subtracting  
the vacuum of the density operator, and not the vacuum of the field. With respect to the known classical
free-streaming solution, we have found quantum corrections related to the difference between the vacuum of
the density operator and the Minkowski vacuum. These corrections are numerically relevant for an early 
decoupling of the field, that is if $m \tau = {\cal O}(1)$, where $\tau$ is the hyperbolic time; in this case
they survive at late times and affect the relation between energy density and pressure as compared to the
classical free-streaming case.

\section*{Acknowledgments}

D.R.\ would like to express his gratitude to D.H.R.\ for his kind hospitality and support at the Institut f\"ur Theoretische Physik of Goethe University Frankfurt am Main during the course of this work.
The work of L.T.\ and D.H.R. was supported by the Deutsche Forschungsgemeinschaft (DFG, German Research Foundation) through the CRC-TR 211 ``Strong-interaction matter under extreme conditions'', project number 315477589 TRR 211.


\appendix

\section{Free streaming in Milne coordinates}\label{freestreaming}

The collisionless Boltzmann equation in classical relativistic kinetic theory reads:
\begin{equation}\label{free}
    p\cdot \partial f(x,{\bf p}) =0\;,
\end{equation}
and its explicit solution in Cartesian coordinates is:
\begin{equation}
    f(x,{\bf p}) = f_0\left({\bf x}-\frac{t-t_0}{\varepsilon}{\bf p},{\bf p}\right)\;,
\end{equation}
where $f_0({\bf x},{\bf p}) = f(t_0,{\bf x};{\bf p})$ is the initial condition in a 
generic inertial reference frame, and $\varepsilon=\sqrt{m^2+p^2}$ is the (on-shell) energy.

In longitudinal boost-invariant symmetry, the initial condition is given at some 
Milne time $\tau_0$ rather than a time $t_0$ in Cartesian coordinates.
Nevertheless, there is a very simple solution in this case, too. Since the distribution function
is a scalar, it must be invariant under the symmetry transformations at stake, that are 
longitudinal boosts as well as rotations and translations in the transverse plane. Hence, it depends 
only on the independent scalars that may be formed with combinations of space-time and momentum 
vector which are invariant under the group of transformations ${\rm IO}(2) \otimes {\rm SO}(1,1)$.
These scalars are:
\begin{align}
    \tau = \sqrt{t^2-z^2}\;, \qquad p_{\rm T} = \sqrt{p_x^2 + p_y^2}\;, \qquad w = z \varepsilon -t p^z\;.
\end{align}
The last variable can be shown to be equivalent to the covariant component $p_\eta$ of the 
four-momentum vector in Milne coordinates. Indeed, there is a fourth invariant scalar:
\begin{equation}
    v = t \varepsilon - z p^z = \tau\sqrt{m^2 + p_{\rm T}^2 +\frac{w^2}{\tau^2}}\;,
\end{equation}
but it is redundant because of the on-shell condition (and positivity) of the energy and because 
$t>|z|$ in the future light cone. The reflection invariance (see Sec.~\ref{symmetry}) makes $f$
dependent on the square of $w$ rather than just $w$. By utilizing these arguments, Eq.~\eqref{free} becomes:
\begin{equation}
    \frac{v}{\tau} \frac{\partial}{\partial \tau} f (\tau,p_{\rm T},w^2)=0\;, 
\end{equation}
since the contribution in the partial derivatives with respect to $w$ cancels out. The free-streaming solution is then very simple, a constant in $\tau$: 
$$
 f(\tau,p_{\rm T},w^2) = f(\tau_0,p_{\rm T},w) \equiv f_0(p_{\rm T},w^2) \;.
$$ 

We are now in a position to calculate the free-streaming solution for the stress-energy 
tensor from its classical kinetic definition:
\begin{equation}
    T^{\mu\nu}=\frac{1}{(2\pi)^3}\int\frac{\di^3{\rm p}}{\varepsilon} p^\mu p^\nu f \Rightarrow \begin{cases}
     {\cal E} = u_\mu u_\nu T^{\mu\nu} = \frac{1}{(2\pi)^3} \int\frac{\di^3{\rm p}}{\varepsilon} \frac{v^2}{\tau^2} f\;, \\
     {\cal P}_{\rm T}=\frac{1}{2}\left( \hat{i}_\mu \hat{i}_\nu + \hat{j}_\mu \hat{j}_\nu  \right)T^{\mu\nu} = \frac{1}{(2\pi)^3} \int\frac{\di^3{\rm p}}{\varepsilon} 
     \frac{p_{\rm T}^2}{2} f\;, \\
     {\cal P}_{\rm L} = \hat{\eta}_\mu \hat{\eta}_\nu T^{\mu\nu} = \frac{1}{(2\pi)^3} \int\frac{\di^3{\rm p}}{\varepsilon} \frac{w^2}{\tau^2} f\;,
    \end{cases}
\end{equation}
and changing the integration variables:
\begin{equation}
    w= z \varepsilon - t p^z \quad \Rightarrow
\quad \di w = \left|-\frac{v}{\varepsilon} \right| \di p^z \quad \Rightarrow \quad \frac{\di p^z}{\varepsilon}= \frac{\di w}{v}\;,
\end{equation}
one obtains:
\begin{equation}\label{freestene}
    {\cal E} = \frac{1}{(2\pi)^3} \int \di^2{\rm p}_{\rm T} \frac{\di w}{v} \; \frac{v^2}{\tau^2}\; f_0(p_{\rm T},w^2) = \frac{1}{(2\pi)^3 \tau} \int \di^2{\rm p}_{\rm T}\,\di w  \; \sqrt{m_{\rm T}^2 + \frac{w^2}{\tau^2}}\; f_0(p_{\rm T},w^2)\;,
\end{equation}
and
\begin{align}\label{freestpre}
    {\cal P}_{\rm T} &= \frac{1}{(2\pi)^3} \int \di^2{\rm p}_{\rm T} \frac{\di w}{v} \; 
    \frac{p_{\rm T}^2}{2}\; f_0(p_{\rm T},w^2) = \frac{1}{(2\pi)^3  \tau} \int 
    \frac{\di^2{\rm p}_{\rm T}\,\di w}{\sqrt{m_{\rm T}^2 + w^2/\tau^2}} \; \frac{p_{\rm T}^2}{2}\; f_0(p_{\rm T},w^2)\;,  \\ \nonumber
    {\cal P}_{\rm L} &= \frac{1}{(2\pi)^3} \int \di^2{\rm p}_{\rm T} \frac{\di w}{v} \; \frac{w^2}{\tau^2}\; f_0(p_{\rm T},w^2) = \frac{1}{(2\pi)^3 \tau} \int \frac{\di^2{\rm p}_{\rm T}\,\di w}{\sqrt{m_{\rm T}^2 + w^2/\tau^2}} \; \frac{w^2}{\tau^2}\; f_0(p_{\rm T},w^2)\;.
\end{align}
The change of variable introduces  an explicit dependence on the proper time in the integral.

Equations \eqref{freestene} and \eqref{freestpre} are the classical relativistic expressions for the
energy density and pressures of a free-streaming gas and coincide with the leading terms obtained in 
Sec.~\ref{setensor} with the substitution $w\to \mu$ and with the initial distribution equal to the 
local equilibrium Bose-Einstein distribution function $f_0=n_{\rm B}^0$.

\section{Minkowski vacuum expectation values}\label{minkoswkivev}

In order to calculate the scalars $\Gamma_{\gamma}(\tau)_M$ of the stress-energy tensor in the Minkowski 
vacuum, we take advantage of it being annihilated by all the $\wh{b}_{\sf p}$'s as it is clear from Eq.~\eqref{atob}. Hence, the only product of $\wh{b}_{\sf p}$ and $\wh{b}_{\sf p}^{\dagger}$ with non-vanishing 
expectation value with respect to $|0_M\rangle$ is $\hat{b}_{\sf p}\hat{b}^{\dagger}_{{\sf p}'}$, and using the 
commutation relations \eqref{commrel}:
\begin{equation}\label{bmink}
	\langle 0_M| \wh{b}_{\sf p}\wh{b}^{\dagger}_{{\sf p}'}|0_M\rangle=
	\langle 0_M| \wh{b}^{\dagger}_{{\sf p}'}\wh{b}_{\sf p}|0_M\rangle + \langle 0_M| [\wh{b}_{\sf p},\wh{b}^{\dagger}_{{\sf p}'}]|0_M\rangle=
	\delta^2(\pT-\pT')\delta(\mu-\mu')\;.
\end{equation}
We can now replace these VEV's to obtain $\Gamma_{\gamma}(\tau)_M$ in the stress-energy tensor 
expression contracted with suitable vectors. For instance, for the energy density, we can use Eq.\ 
\eqref{rhole} by simply replacing the local equilibrium expectation values with those in the 
Minkowski vacuum and obtain:
\begin{equation}
	{\cal E}(\tau)_M\equiv \langle 0_M|\wh{T}^{\mu \nu}u_{\mu}u_{\nu}|0_M\rangle=
	\int \frac{{\rm d^2p_T}\,{\rm d}\mu}{4(4\pi)^2}\left(|\partial_{\tau}h|^2+\omega^2|h|^2\right)=
	\int \frac{{\rm d^2p_T}\,{\rm d}\mu}{(2\pi)^3\tau}\omega \frac{K}{2}\;.
\end{equation}
Similarly, for the pressures, one finds:
\begin{equation}\label{gammamink}
	\Gamma_{\gamma}(\tau)_M=\int \frac{{\rm d^2p_T}\,{\rm d}\mu}{(2\pi)^3\tau}\omega \frac{K_{\gamma}}{2}\;.
\end{equation}
%

\section{Asymptotics}\label{asymptotics}

It is interesting to study the behaviour of the stress-energy tensor and related quantities for
late times $\tau$. With
\begin{equation}\label{hankelint}
    h(\tau)=-i{\rm e}^{\frac{\pi}{2}\mu}{\rm H}^{(2)}_{i\mu}(\mT \tau)\;,
\end{equation}
one can make use of the asymptotic expansion for large arguments~\cite{Gradshteyn}
\begin{equation}\label{large_x}
    {\rm H}^{(2)}_{\nu}(x)\sim\sqrt{\frac{2}{\pi x}}\e^{-\ii \left( x-\frac{\pi}{2}\nu -\frac{\pi}{4} \right)}\sum_n\frac{1}{(2ix)^n} \frac{\Gamma(\nu+1/2 +n)}{n!\Gamma(\nu+1/2-n)}\;,
\end{equation}
which is valid for ${\rm Re}(\nu)>-1/2$ and $|{\rm arg}(x)|<\pi$. Making use of the property 
$z\Gamma(z) = \Gamma(z+1)$, substituting $x=m_{\rm T}\tau$ and $\nu=i\mu$, and plugging this
into Eq.~\eqref{hankelint} we get:
$$
    h(\tau) \sim \sqrt{\frac{-2i}{\pi m_{\rm T}\tau}}\e^{-\ii m_{\rm T}\tau}\sum_n
\frac{1}{(2 \ii  m_{\rm T}\tau)^n} \frac{\left(i\mu +\frac{1}{2} -n \right)^{(2n)}}{n!}\;,
$$
valid for large $m_{\rm T}\tau$. Similarly, using the exact relation~\cite{Gradshteyn}
$$
   z \partial_z {\rm H}^{(2)}_{\nu}(z) =\nu {\rm H}^{(2)}_{\nu}(z) -z {\rm H}^{(2)}_{\nu+1}(z)\;,
$$
along with the expansion~(\ref{large_x}), one obtains the expansion for the proper-time derivative 
$\partial_\tau h$:
$$
    \partial_\tau h(\tau) \sim -\ii m_{\rm T}\sqrt{\frac{-2 \ii}{\pi m_{\rm T}\tau}} 
    \e^{-\ii m_{\rm T}\tau}\left[1+ \sum_{n>0} \frac{1}{(2\ii m_{\rm T}\tau)^{n}}
    \left(-2\ii\mu \frac{\left(i\mu +\frac{3}{2} -n \right)^{(2n-2)}}{(n-1)!} + 
    \frac{\left(\ii \mu +\frac{3}{2} -n \right)^{(2n)}}{n!}\right) \right]\;.
$$
In particular, retaining the terms up to first order (i.e., next-to-leading order) in $m_{\rm T}\tau$ 
we get:
\begin{align*}
 h(\tau) &\simeq \sqrt{\frac{-2i}{\pi m_{\rm T}\tau}}e^{-\ii m_{\rm T}\tau}
 \left[ 1 -\frac{\ii}{2m_{\rm T}\tau} \left( \ii \mu -\frac{1}{2}\right)\left( \ii \mu +\frac{1}{2}\right)\right] = \sqrt{\frac{-2\ii }{\pi m_{\rm T}\tau}}\e^{-\ii m_{\rm T}\tau}\left(1 +\ii \frac{1+4\mu^2}{8m_{\rm T}\tau} \right)\;,  
 \\
 \partial_{\tau} h(\tau) &\simeq -\ii m_{\rm T}\sqrt{\frac{-2\ii}{\pi m_{\rm T}\tau}}
 \e^{-\ii m_{\rm T}\tau}\left(1-\ii \frac{3-4\mu^2}{8m_{\rm T}\tau} \right)\;.
\end{align*}
Feeding the above expansions into the definitions~\eqref{kappa2} and~\eqref{lambda2}, 
\begin{equation}\label{kappa3}
    K_\gamma\simeq\frac{m_{\rm T}^2 +\gamma}{2 m_{\rm T}\omega}\;,
\end{equation}
\begin{equation}\label{lambda3}
    \Lambda_\gamma\simeq\frac{1}{2m_{\rm T} \omega}\e^{-2 \ii m_{\rm T}\tau}
    \left[-m_{\rm T}^2\left( 1 -\ii \frac{3-4\mu^2}{4m_{\rm T}\tau} \right) + 
    \gamma \left( 1 +\ii \frac{1+4\mu^2}{4m_{\rm T}\tau} \right) \right]\;.
\end{equation}
The rest is of the order of $1/[m_{\rm T}\omega(m_{\rm T}\tau)^2]$ for $K_\gamma$ and
$\exp( -2 \ii m_{\rm T}\tau)/[m_{\rm T}\omega(m_{\rm T}\tau)^2]$ for $\Lambda_\gamma$.

Equations \eqref{kappa3} and \eqref{lambda3} are very useful to study the large $p_{\rm T}$ 
(hence, large $m_{\rm T}$) behaviour as well as the long-time behaviour. For large $p_{\rm T}$, 
Eq.~\eqref{kappa3} implies that $K \to 1$, hence to leading order  $K-1$ is simply zero.
However, from Eq.\ \eqref{lambda3} and the exact relation~(\ref{reltheta}) one can obtain the terms up to
second order. Indeed, for $\gamma=\omega^2$, in the large $m_{\rm T}$ limit:
\begin{equation}
    \Lambda \simeq \frac{1}{2m_{\rm T} \tau} \e^{-2\ii m_{\rm T}\tau}\;,
\end{equation}
hence: 
\begin{equation}
  |\Lambda| = \sinh\Theta \simeq \Theta = \frac{1}{2m_{\rm T}\tau}\;, 
\end{equation}
and in the limit of large $m_{\rm T}$
\begin{equation}
    K-1 = \cosh \Theta \simeq\frac{1}{2}\Theta^2 \simeq \frac{1}{8m_{\rm T}^2\tau^2}\;.
\end{equation}

Similarly, the leading expressions at late time $\tau\to\infty$ can be derived. By using the asymptotic 
expansions \eqref{kappa3} and~\eqref{lambda3} and expanding $\omega(\tau)$ and $\gamma(\tau)$ for 
large $\tau$ one obtains:
\begin{align}
    K_\gamma &\simeq \frac{m_{\rm T}^2 +\tilde \gamma}{2 m_{\rm T}^2}\;, \\ \nonumber 
    \Lambda_\gamma &\simeq \frac{1}{2m_{\rm T}^2 }\e^{-2 \ii m_{\rm T}\tau}
    \left[\tilde\gamma - m_{\rm T}^2 +\ii \frac{m_{\rm T}^2(3-4\mu^2)+\tilde 
    \gamma(1+4\mu^2)}{4m_{\rm T}\tau}\right]\;,
\end{align}
at first order in $1/\tau$, with $\tilde\gamma$:
\begin{equation}
    \tilde \gamma = \lim_{\tau\to\infty} \gamma =\begin{cases}
        m_{\rm T}^2\;, & \mbox{for }{\cal E}\;, \\
        -m^2\;, & \mbox{for } {\cal P}_{\rm T}\;,\\
        -m_{\rm T}^2\;, &\mbox{for } {\cal P}_{\rm L}\;.
    \end{cases}
\end{equation}

It is important to note that, except for the energy density and only at the leading order, the 
$\Lambda_\gamma$'s have a rapidly oscillating phase that prevents a proper limit in the function 
domain. However, they converge in the distribution domain, which is fine since they have to be 
integrated. In fact the limits:
$$
    \lim_{\tau\to\infty} \sin(2m_{\rm T}\tau)\;, \qquad \lim_{\tau\to\infty} \cos(2m_{\rm T}\tau)
$$
are proportional to Dirac deltas. We make use of the formula for the delta families
$$
 \delta(x)=\lim_{\epsilon\to 0} \frac{1}{\epsilon}f(x/\epsilon)\;,
$$
for any $f$ function normalized to $1$ and set $\epsilon=1/\tau$ in the former case and 
$\epsilon = 1/\sqrt{\tau}$ in the latter:
\begin{equation}
 \begin{split}
 & \lim_{\epsilon\to 0} \frac{1}{\pi}\frac{\sin(x/\epsilon)}{x} = \delta(x)\quad  \Rightarrow \quad \sin(2m_{\rm T}\tau)\;\stackrel{\tau\to \infty}{\longrightarrow}\;2\pi m_{\rm T}\; \delta(2m_{\rm T})\;, \\
  &\lim_{\epsilon\to 0} \frac{1}{\epsilon \sqrt{\pi}}\cos\left(\frac{x^2}{\epsilon^2}\right) = \delta(x)\quad  \Rightarrow \quad  \cos(2m_{\rm T}\tau)\;\stackrel{\tau\to \infty}{\longrightarrow}\; \sqrt{\frac{\pi}{\tau}}\, \delta(\sqrt{2m_{\rm T}})\;.
   \end{split}
\end{equation}
In both cases the Dirac delta is outside of the domain of integration, and all these integrals are vanishing in the long proper-time limit.

\end{document}